\documentclass[preprint,12pt,review]{elsarticle}
\usepackage{amssymb}
\usepackage{amsmath}
\usepackage{multicol}
\usepackage{multirow}
\usepackage{array}
\usepackage[table]{xcolor}
\usepackage{hyperref}
\usepackage{longtable}

\journal{}

\begin{document}

\begin{frontmatter}

\title{Modelling the impact of repeat asymptomatic testing policies for staff on SARS-CoV-2 transmission potential}

\author[MAN]{Carl A. Whitfield
\footnote{Corresponding author: carl.whitfield@manchester.ac.uk}}
\author[MAN]{University of Manchester COVID-19 Modelling Group
\footnote{This recognises the equal contributions of the following authors: Jacob Curran-Sebastian, Rajenki Das, Elizabeth Fearon, Martyn Fyles, Yang Han, Thomas A. House, Hugo Lewkowicz, Christopher E. Overton, Xiaoxi Pang, Lorenzo Pellis, Heather Riley, Francesca Scarabel, Helena B. Stage, Bindu Vekaria, Feng Xu, Jingsi Xu, Luke Webb.}}
\author[MAN]{Ian Hall}

\affiliation[MAN]{Department of Mathematics, University of Manchester,United Kingdom}

\begin{abstract}
Repeat asymptomatic testing in order to identify and quarantine infectious individuals has become a widely-used intervention to control SARS-CoV-2 transmission. In some workplaces, and in particular health and social care settings with vulnerable patients, regular asymptomatic testing has been deployed to staff to reduce the likelihood of workplace outbreaks. We have developed a model based on data available in the literature to predict the potential impact of repeat asymptomatic testing on SARS-CoV-2 transmission. The results highlight features that are important to consider when modelling testing interventions, including population heterogeneity of infectiousness and correlation with test-positive probability, as well as adherence behaviours in response to policy. Furthermore, the model based on the reduction in transmission potential presented here can be used to parameterise existing epidemiological models without them having to explicitly simulate the testing process. Overall, we find that even with different model paramterisations, in theory, regular asymptomatic testing is likely to be a highly effective measure to reduce transmission in workplaces, subject to adherence. This manuscript was submitted as part of a theme issue on "Modelling COVID-19 and Preparedness for Future Pandemics". 
\end{abstract}


\begin{highlights}
\item Model of SARS-CoV-2 test sensitivity and infectiousness based on data freely available in the literature
\item Simple, efficient algorithm for simulating testing in a heterogeneous population
\item Regular lateral flow tests have a similar impact on transmission to PCR tests 
\item Adherence behaviour is crucial to actual testing impact
\item Regular testing reduces the size and likelihood of outbreaks in closed populations.
\end{highlights}


\end{frontmatter}



\section{Introduction}
\label{sec:intro}

In the early stages of the COVID-19 pandemic, many countries had limited capacity for SARS-CoV-2 diagnostic testing, and so a large proportion of asymptomatic or `mild' infections were not being detected. At this stage in the UK, testing was primarily being used in hospitals for patient triage and quarantine measures. By late 2020, many western countries had greatly increased their capacity to perform polymerase chain reaction (PCR) tests, which sensitively detect SARS-CoV-2 RNA from swab samples taken of the nose and/or throat. Importantly, several types of antigen test in the form of lateral flow devices (LFDs) came to market, promising to detect infection rapidly (within 30 mins of the swab) and inexpensively in comparison to PCR. In the UK, many of these devices underwent extensive evaluation of their sensitivity and specificity in lab and real-world settings \cite{peto_2021}, and several were made freely available to the general public. 

Nonetheless, there was significant concern surrounding the sensitivity of LFD tests \cite{deeks_2020,wise_2020}, their sensitivity was observed to be low relative to PCR in early pilot studies \cite{dinnes2020,ferguson_2020,garcia-finana_2021,mina_2021} and so they were more likely to miss true positives. Furthermore, in some cases, extremely low values for their specificity were reported \cite{armstrong_2020,kanji_2021,gans_2022}, suggesting that false positives may be common, although this later appeared not to be the case for the devices that were systematically tested and rolled out to the wider public in the UK \cite{peto_2021,wolf_2021}. Models of SARS-CoV-2 testing and observational studies predicted that regular asymptomatic testing and contact-tracing could significantly reduce transmission rates in the population \cite{hellewell_2021,ferretti_2021,fyles_2021,lee_2022}. Furthermore, specific studies on hospitals \cite{evans_2021}, care-homes \cite{rosello_2022}, and schools \cite{asgary_2021,leng_2022} have demonstrated the impact that regular LFD testing can have and has had on reducing transmission in these vital settings. However, studies have also highlighted the potential pitfalls and inefficiencies of such policies \cite{tulloch_2021}, in particular around the factors affecting adherence to these policies. Testing can only reduce transmission if it results in some contact reduction or mitigation behaviour in those who are infectious. Therefore, studies have shown the importance of ensuring that isolation policies in workplaces are coupled with measures to support isolation, such as paid sick-leave \cite{ahmed_2020,patel_2021,daniels_2021}.

The focus of this paper is modelling the potential impact of testing in workplaces and in particular its effect on reducing the transmission of SARS-CoV-2 in these settings. We consider several of the confounding factors already discussed, including test sensitivity and adherence to policy, as well as highlighting some other important features including the impact of population heterogeneity. These questions are addressed using data on the within-host dynamics of SARS-CoV-2 viral load as well as data on test sensitivity and infectiousness that is available in the literature. The model we present is generic, and similar to those used in \cite{hellewell_2021,quilty_2021,ferretti_2021,fyles_2021}, in order to be applicable to a wide range of settings and scenarios. However, we also extend some of these findings to the care home setting, to understand its implications for evaluating and comparing potential testing policies in care homes. To give context, the results presented in section \ref{sec:main_results}, were used to inform policy advice for staff testing in social care from the Social Care Working Group (a sub-committee of the Scientific Advisory Group for Emergencies - SAGE) which was reported to Department for Health and Social Care (DHSC) and SAGE in the UK \cite{scwg_report_2021}. This paper takes a more detailed look into the predictions of this model and its implications for implementing testing policies in workplaces in general. 

\section{Methods}
\label{sec:methods}

We focus on quantifying the effects of testing and isolation on a single individual, for which we use the concept of an infected person's ``infectious potential''. This was then extended to investigate the impact of testing in a generic workplace setting on transmission of SARS-CoV-2 making some basic assumptions regarding contact and shift patterns. 

Table \ref{tab:symbols} provides a list of all the symbols we use to represent the parameters and variables in this section, alongside a description of each.

\begin{longtable}{| m{2.5cm}<{\centering} | m{10cm} |}
\hline Symbol & Description  \\ \hline
\multicolumn{2}{|c|}{\textbf{(a) Transmission modelling}} \\ \hline
$R_0$, $R_{\rm ind}^{k}$ & Basic reproduction number and expected reproduction number of individual $k$ respectively. \\ \hline
$c_0$, $c_k(t)$ & Basic contact rate and contact rate of individual $k$ respectively. \\ \hline
$p_k(t)$ & Probability of a contact between infectious individual $k$ and a susceptible individual at time $t$ resulting in an infection. \\ \hline
$\beta_0$, $J_k(t)$ & Baseline probability of a contact resulting in an infection and relative infectiousness of individual $k$ at time $t$. \\ \hline
$X$ & Fractional reduction in contact rate due to isolation ($X=1$ used throughout). \\ \hline
$t_{\rm isol}^{(k)}$, $\tau_{\rm isol}$ & The time since infection when the individual begins isolation and the duration of the isolation period respectively. \\ \hline
$\tau_{\inf}^{(k)}$ & Duration of infection of individual $k$, defined as infectious when nasal viral load is detectable by PCR. \\ \hline 
\multicolumn{2}{|c|}{\textbf{(b) Viral load models}} \\ \hline
$V_k(t)$ & Nasal viral load of individual $k$ at time $t$ since infection. \\ \hline
$V_p^{(k)}$, $t_p^{(k)}$ & Peak viral load value of individual $k$ and the time since infection it occurs respectively. \\ \hline
$r_k$, $d_k$ & Exponential rate of viral growth and decay respectively in individual $k$. \\ \hline
$V_{\rm lod}$, $\Delta$Ct, $\omega_p$ $\omega_r$ & Viral load parameters in reference \cite{kissler_2021}\\ \hline
\multicolumn{2}{|c|}{\textbf{(c) Infectiousness model}} \\ \hline
$J_p^{(k)}$, $h_k$ & Theoretical maximum infectiousness (at high viral load) and steepness of Hill function relating infectiousness to viral load respectively for individual $k$. \\ \hline
$K_m$ & Threshold parameter for Hill function of viral load vs. infectiousness (same for all individuals). \\ \hline
IP$_k$ & Infectious potential of individual $k$, a values of 1 indicates the same overall infectiousness as the population mean without any isolation. \\ \hline 
$\Delta$IP & Relative reduction in IP compared to some baseline case. \\ \hline
\multicolumn{2}{|c|}{\textbf{(d) Testing models}} \\ \hline
$P_{\rm PCR}(V)$, $P_{\rm LFDh}(V)$, $P_{\rm LFDl}(V)$ & Probability of positive test result given a viral load of $V$ for PCR, high-sensitivity LFD, or low-sensitivity LFD respectively.\\ \hline
$P_{\rm max}$, $s_p$, $V_{50}^{(p)}$ & PCR sensitivity parameters: maximum sensitivity, slope of logistic function, and threshold viral load for logistic function respectively. \\
\hline
$\lambda$, $s_l$, $V_{50}^{(l)}$ & LFD ``high sensitivity'' parameters (relative to PCR sensitivity): maximum sensitivity, slope of logistic function, and threshold viral load for logistic function respectively. \\
\hline
$P_{\rm not}$ $P_{\rm miss}$ & Parameters of adherence: the proportion of people who do no tests and the proportion of tests missed by those who do test respectively. \\ \hline
$A$ & Composite adherence parameter (i.e. the proportion of tests that get performed, $A = (1 - P_{\rm not})(1 - P_{\rm miss})$). \\ \hline
$Z$, $p$, $\tau_{\rm pos}$ & Parameters of simple testing model: fractional impact of isolation on infectiousness, test sensitivity in window of opportunity, and duration of window of opportunity respectively. \\ \hline
$\tau$ & Time between tests for a given regular mass testing policy. \\ \hline
\multicolumn{2}{|c|}{\textbf{(e) Workplace models}} \\ \hline
$W_k(t)$ & Shift pattern indicator ($=1$ when individual $k$ is at work, and $=0$ otherwise).\\ \hline
$N_s$, $N_r$ & Number of employees in a model workplace and the number of residents in the care-home model respectively. \\ \hline
$f_s$, $p_c$ & Fraction of days staff spend on-shift (9/14 used here) and probability of them making a contact during a shift with a particular co-worker who is also in work that day respectively. \\ \hline
$a$, $b$ & Fractional contact probabilities (relative to $p_c$) in model care-home between staff and between residents respectively. \\ \hline
\caption{Symbols used in this paper to represent various mathematical variables and parameters and their interpretation, broken down by category.}
\label{tab:symbols}
\end{longtable}

\subsection{Infectious and Transmission potential}
\label{sec:inf_potential}

Without loss of generality, we define the time an individual $k$ gets infected as $t = 0$. Assuming non-repeating contacts at rate $c_k(t)$ then the expected number of transmission events from that infected individual (or their \emph{Transmission potential} \cite{quilty_2021}) is
\begin{align}
    \label{Rind} R_{\rm ind}^{(k)} = \int_0^\infty c_k(t) p_k(t) \textrm{d} t.
\end{align}

Initially, we aim to gain generic insights into how testing can impact on transmission, and so we consider one other simplifying assumption, that the contact rate $c_k(t)$ can be described by a simple step function, such that the contact rate is reduced by a factor $X$ when an individual self-isolates
\begin{align}
\label{contact_rate} c_k(t) =  \begin{cases}
c_0 \qquad & \textrm{if} \; t < t_{\rm isol}^{(k)} \; \textrm{or} \; t \geq t_{\rm isol}^{(k)} + \tau_{\rm isol}\\
(1-X)c_0 \qquad & \textrm{if} \; t_{\rm isol}^{(k)} \leq t < t_{\rm isol}^{(k)} + \tau_{\rm isol},
\end{cases}
\end{align}
where the parameters are as defined in table \ref{tab:symbols}(a).

Finally, we suppose that the probability of transmission per contact event is $p_k(t) = \beta_0 J_k(t)$ where $\beta_0$ is a constant. We define the (arbitrary) scaling of the infectiousness $J(t)$ by setting $\langle \int_0^{\infty} J(t) {\rm d} t \rangle = \langle{\tau}_{\rm inf}\rangle $, where $\langle \cdot \rangle$ denotes a population average such that $\langle x \rangle \equiv \sum_{k=1}^{N}x_k/N$. The parameter $\langle\tau_{\rm inf}\rangle$ is the average period for which a person can test positive via PCR (i.e. how long they have a detectable COVID infection). Therefore, ignoring isolation, if the contact rate $c_0$ is the same for all individuals (note that this is not a necessary assumption, variations in contact rate between individuals can be absorbed into the infectiousness by a simple scaling factor) then the population baseline reproduction number without any isolation will be $R_0 = \beta_0 c_0 \langle\tau_{\rm inf}\rangle$. Then we can rewrite equation \eqref{Rind}, the individual reproduction number for individual $k$, as
\begin{align}
    \label{Rnum} R_{\rm ind}^{(k)} = \frac{R_0}{\langle \tau_{\rm inf} \rangle} \int_0^{\infty} J_k(t)\left[1 -  X I(t;t_{\rm isol}^{(k)},t_{\rm isol}^{(k)} + \tau_{\rm isol}^{(k)}) \right] \, \mathrm{d} t \equiv R_0 \textrm{IP}_k
\end{align}
where $I(x;a,b) = H(x-x_1) - H(x-x_2)$ is an indicator function for the range $x_1 < x < x_2$ such that $H(x)$ is the Heaviside step function. The quantity IP$_k$ we define as the individual's \emph{infectious potential} and under these simplifying assumptions is proportional to the individual's reproduction number in a fully susceptible population. More generally IP is the relative infectiousness of an individual (omitting any isolation) integrated over the infectious period, and so still has epidemiological relevance beyond the case of a fully susceptible population.

Throughout this paper, we will use the relative reduction in IP vs. some baseline scenario (generally a scenario with no testing), to measure the impact of testing regimes, such that
\begin{align}
    \label{DeltaIP} \Delta \textrm{IP} =  1 - \frac{\langle\textrm{IP}\rangle}{\langle\textrm{IP}_0\rangle},
\end{align}
where $\langle\textrm{IP}_0\rangle$ is the population average IP for the baseline scenario. 

\subsection{Model of viral-load, infectiousness and test positive probability}
\label{fig:parameterisations}

We use a RNA viral-load based model of infectiousness and test positive probability, similar to those used in \cite{quilty_2021,ferretti_2021} to calculate the reduction in IP for different testing and isolation behaviour. Individual viral-load trajectories $V_k(t)$ are generated to represent the concentration of RNA (in copies/ml) that should be measured in PCR testing of a swab of the nasal cavity. In turn these are used to calculate an infected individual's infectiousness $J(V(t))$ and probability of testing positive $P(V(t))$ over time. 

We assume, as in \cite{kissler_2021}, that the viral load trajectory can be described by the following piecewise exponential (PE) model
\begin{align}
    \label{VLmodel}
        V_k(t) = \begin{cases}
        V_p \exp[r(t_p - t)] \quad & \textrm{if} \; t \le t_p\\
        V_p \exp[-d(t - t_p)] \quad & \textrm{if} \; t > t_p\\
        \end{cases},
\end{align}
where the parameters are as defined in table \ref{tab:symbols}(b). We use two different datasets to parameterise the PE model, which are laid out in the following sections.

\subsubsection{Parameterisation of RNA viral-load model}
\label{sec:viral_load_param}

\textbf{Ke et al. (2021) data:}

The PCR-measured viral-load trajectories (in RNA copies/ml) are generated at random based on the mechanistic model fits in \cite{ke_2021}. That dataset consists of the results of daily PCR and virus culture tests to quantify RNA viral-load (in RNA copies/ml) and \emph{infectious} viral load, in arbitrary units akin to plaque-forming units (PFUs) respectively. There were 56 participants who had been infected by different variants of SARS-CoV-2 (up to the delta strain, as this dataset precedes the emergence of omicron). In order to use this data, we first simulated the ``refractory cell model'' (RCM) described in \cite{ke_2021} for all 56 individual parameter sets given in the supplementary information of that article. Note that in \cite{ke_2021} these were based on nasal swabs, we did not use the data from throat swabs. 

We found that, at long times, the RCM would show an (unrealistic) second growth stage of the virus. To remove this spurious behaviour from these trajectories, we fitted the parameters of the PE model (outlined in the previous section) to the data around the peak viral load. To do this, we truncated the data to only consist of only viral load values around the first peak that was above a threshold of $V_{\rm thresh} = \max{V}^{0.5}$ (i.e. half of the maximum viral load on a log-scale, measured empirically from the generated trajectory). The data for each trajectory was then truncated between two points to avoid fitting this spurious behaviour. The first point was when the viral load first surpassed $V_{\rm thresh}$. The second point was either when viral load fell below $V_{\rm thresh}$, or the data reached a second turning point (a minimum) -- whichever occurred first. We then used a simple least-squares fit on the log-scale to fit the PE model to this truncated data. In order to set realistic initial values for the non-linear least-squares fitting method, we estimated the peak viral load and time by taking the first maximum of the viral load trajectory and the time it occurred. Then estimated the growth and decay rates by simply taking the slope of straight lines connecting the viral load at the start and end of the truncated data to this peak value.

\begin{figure}[ht]
\centering
\includegraphics[width=\textwidth]{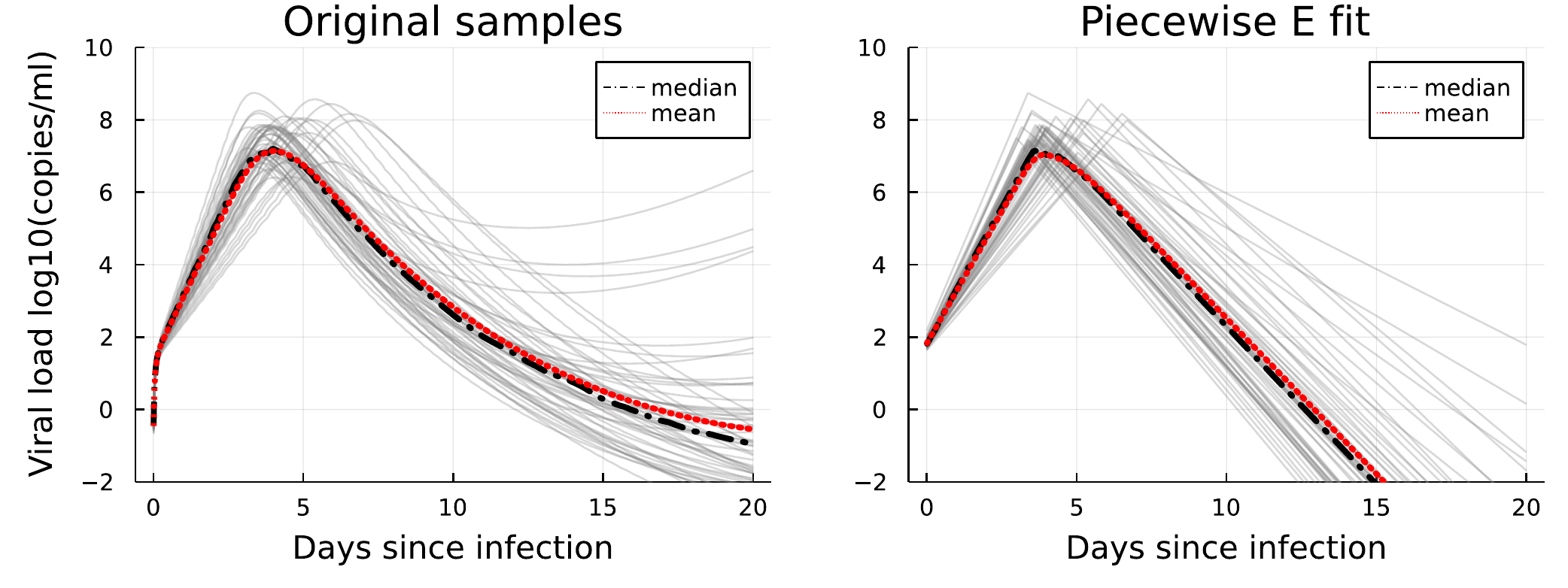}
\caption{Left: Spaghetti plot of the 56 RCM model paramterisations given in \cite{ke_2021}. Right: the same plot but showing the PE model parameterisations fitted in this paper. Note that the point-wise median and mean profiles shown here were computed on the scale of $\log_{10}$copies/ml.}
\label{fig:example_PE_fit}
\end{figure}

Figure \ref{fig:example_PE_fit} shows all of the PE fits against the original RCM model fits. We can see that the PE model captures the dynamics around the peak well (which, for our purposes, is the most important part of the trajectory as it is when individuals are most likely to be infectious and test positive). However, this fitting comes at the expense of losing information about the changing decay rate at longer times (which has a much smaller effect on predicting testing efficacy). 

Supplementary table S1 summarises the mean and covariance of the maximum likelihood multivariate lognormal distribution of the PE model parameters. We use this distribution to generate random parameter sets for the PE model which are used to simulate different individuals. 

\textbf{Kissler et al. (2021) data:}

\begin{figure}[ht]
    \centering
    \includegraphics[width=\textwidth]{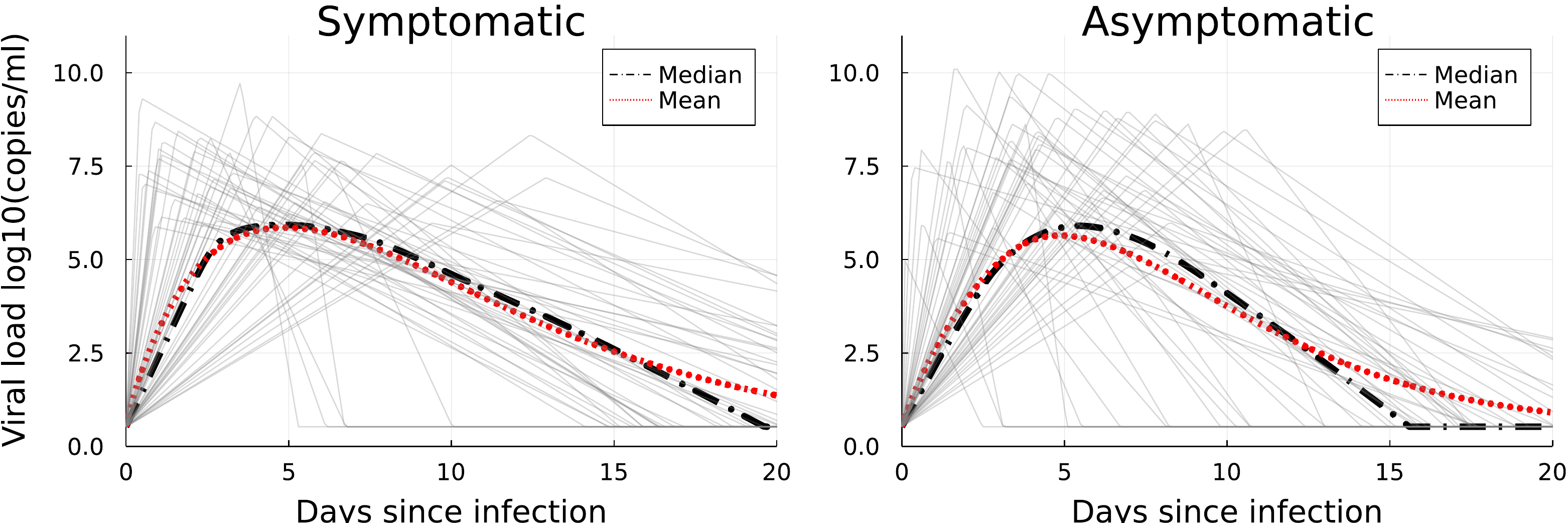}
    \caption{Spaghetti plot of trajectories generated using random samples of the posterior distribution of PE parameters in \cite{kissler_2021}. The point-wise mean and median lines are computed using 10,000 samples. Note that, for the purposes of this plot, the trajectories are truncated so that any values below $V_{\rm lod}$ are set to $V_{\rm lod}$ (so that spuriously small values on the log-scale do not affect the averages. The left graph shows samples from the symptomatic population in that study and right the asymptomatic.}
    \label{fig:kissler_samples}
\end{figure}

The data from \cite{kissler_2021} consists of 46 individuals identified to have ``acute'' SARS-CoV-2 infections while partaking in regular PCR tests. To simulate the data from this paper, we sample the individual-level posteriors (available at \cite{kissler_data_2021}) directly. First, we converted the parameters contained in that dataset ($\Delta Ct$, difference between minimum Ct and the limit of detection (LoD); $\omega_p$, length of growth period from LoD to peak viral-load; and $\omega_r$ length of decay period from peak viral-load to LoD) as follows
\begin{align}
    \log_{10}(V_p) &= \log_{10}(V_{\rm lod}) + \frac{\Delta Ct}{3.60971}, \\
    r &= \log(10)\left[\frac{\log_{10}(V_p) - \log_{10}(V_{\rm lod})}{\omega_p}\right], \\
    d &= \log(10)\left[\frac{\log_{10}(V_p) - \log_{10}(V_{\rm lod})}{\omega_r}\right].
\end{align}
The parameter $V_{\rm lod} = 10^{2.65761}$copies/ml is the viral load corresponding to a Ct-value of 40 and $3.60971$ is the fitted slope between Ct-value and RNA viral-load in $log_{10}$(copies/ml) in that study. Finally, in order to determine the final parameter $t_p$, we used the result from \cite{ferretti_2021} based on the same dataset that the viral load at time of infection is $V_0 = 10^{0.5255}$copies/ml, such that 
\begin{align}
    t_p = \frac{\log V_p - \log V_0}{r}.
\end{align}
We separated the converted datasets into those individuals who were labelled as ``symptomatic'' and those who were not, as there was shown to be statistically significant differences in these populations in \cite{kissler_2021}. Then, to generate a new viral load trajectory a single parameter set is sampled from one of these two datasets, depending on whether the trajectory corresponds to a simulated individual who would develop symptoms or not. Example trajectories for the two cases are shown in figure \ref{fig:kissler_samples}.

\subsubsection{Parameterisation of infectiousness as a function of RNA viral load}
\label{sec:infectiousness_param}

\begin{figure}[ht]
\centering
\includegraphics[width=\textwidth]{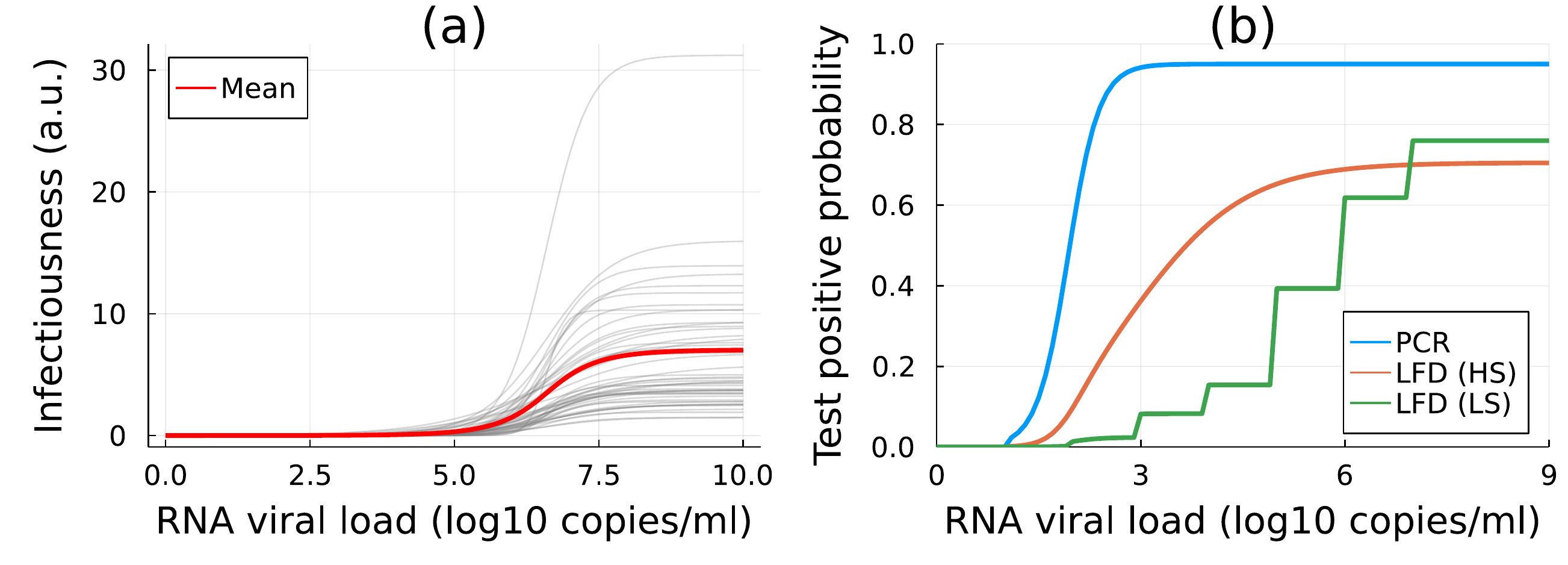}
\caption{(a) The deterministic relationship between infectiousness and RNA viral load used here. Grey lines show individual samples as each individual is assumed to have a different random value of $J_p$ and $h$. The red line shows the population mean. (b) The different relationships between test-positive probability and RNA viral load used in this paper. The blue line shows the assumed PCR test senstivity, while orange  shows the `high' sensitivity (HS) LFD case and green shows the `low' sensitivity (LS) case (note that the peak sensitivity is actually higher in the LS case, but overall sensitivity is lower).}
\label{fig:test_and_inf}
\end{figure}

In order to model infectiousness, we use ``infectious virus shed'' as fitted in \cite{ke_2021} as a proxy. In \cite{ke_2021} infectious virus shed is a Hill function of viral load
\begin{align}
    \label{infectiousness} J_k(V_k) = \frac{J_p V_k^h}{V_k^h + K_m^h}.
\end{align}
We found that neither of the random parameters $J_p$ nor $h$ (as given in \cite{ke_2021}) were significantly correlated to any of the PE model parameters fitted for the same individuals. Therefore, we used the maximum likelihood bivariate lognormal distribution of the random parameters of this model $\{J_p, h\}$, given in Supplementary table S1. These infectiousness parameters are generated independently of the individual's RNA viral-load parameters $\{V_p, t_p, r, d\}$. Examples of $J_k(x)$ are shown in figure \ref{fig:test_and_inf}(a) as well as the mean relationship.

Note that, in \cite{ke_2021}, the magnitude of $J_p$ is given in arbitrary units and so the IP measure we use here is also in arbitrary units. Thus, we present results in terms of a relative reduction in IP ($\Delta$IP), which is independent of the choice of infectiousness units. 

\subsubsection{Parameterisation of test sensitivity as a function of RNA viral load}
\label{sec:test_sens}

The probability of testing positive is also assumed to be deterministically linked to RNA viral load. Note that we assume this because of the data available on LFD test sensitivity as a function of RNA viral load measured by PCR, however several studies have suggested that LFD test sensitivity is actually more closely linked to infectious viral load or culture positive probability \cite{kirby_2021, pekosz_2021, pickering_2021, killingley_2022}. Furthermore, the sensitivity relationships used here imply that the outcomes of subsequent tests are independent. This is likely to be an unrealistic assumption in practice since factors other than viral load may also influence test outcome, and these may vary between individuals.

To model PCR testing, we use a logistic function with a hard-cutoff to account for the cycle-threshold (Ct) cutoff
\begin{align}
P_{\rm PCR}(V_k) = 
\begin{cases}
0 \quad & \textrm{if} \; V_k < V_{\rm cut}\\
P_{\rm max}\left[1 + e^{-s_p(\log_{10} V_k - \log_{10} V_{50}^{(p)}}\right]^{-1} \quad & \textrm{if} \; V_k \geq V_{\rm cut}
\end{cases}.
\end{align}
The parameters $P_{\rm max}$, $V_{50}$ and $s_t$ are extracted by a maximum-likelihood fit of the data on the ``BioFire defense'' PCR test given in \cite{smith_2020}, and are given in Supplementary table S1. Note that we fitted to logistic, normal cumulative distribution function (CDF), and log-logistic functions and chose maximum likelihood fit. 

To model LFD testing, we use two data sources to establish a `low' and 'high' sensitivity scenario. In the `high' sensitivity case we use the phase 3b data collected in \cite{peto_2021}. We fitted logistic, normal CDF and log-logistic models to the data using a maximum likelihood method and chose the most likely fit (which was the logistic model). The phase 3b results in this study came from community testing in people who simultaneously tested positive by PCR. Therefore, the overall LFD sensitivity is given by the logistic function multiplied by the fraction who would test positive by PCR, i.e.
\begin{align}
\label{PCRhigh} P_{\rm LFDh}(V_k) = \lambda P_{\rm PCR}(V_k)\left[1 + e^{-s_l(\log_{10} V_k - \log_{10} V_{50}^{(l)})}\right]^{-1}.
\end{align}
The parameters $s_l$ and $V_{50}^{(l)}$ are determined by maximum likelihood fit, which the relative sensitivity $\lambda$ is included to account the difference in sensitivity between self-testing and testing performed by lab-trained staff \cite{peto_2021}.

In the `low' sensitivity LFD case, we use data from regular LFD and PCR testing in the social care sector in the UK \cite{nhs_test_and_trace_dual-technology_2021}. Since this data is based on positive PCR results, we assume again $P_{\rm LFDl}(V_k) = \theta(V_k)P_{\rm PCR}(V_k)$. The function $\theta(V_k)$ is a stepwise function, and is parameterised in Supplementary table S1. All of the test-positive probability relationships are shown in figure \ref{fig:test_and_inf}(b).

\subsection{Adherence to policy}
\label{sec:adherence_model}

In this paper we consider a number of testing policies which take the form of instructions to employees in the workplace regarding the number of tests to carry out per week. In general, we assume that PCR tests are carried out with 100\% adherence, as these are assumed to be `enforced' (our reference example is PCR testing of hospital and care home staff in the UK, which are carried out at the workplace by trained staff). LFD tests on the other hand are assumed voluntary, as these are carried out at home and reported online. We consider two behavioural parameters to model how individuals may choose to adhere with LFD testing policies, $P_{\rm not}$ and $P_{\rm miss}$ (see table \ref{tab:symbols}(d)). In sections \ref{sec:adherence_results} we explicitly model two behavioural extremes, 
\begin{itemize}
    \item ``All-or-nothing'': $P_{\rm not} = 1 - A$ and $P_{\rm miss} = 0$, i.e. a fixed fraction of people complete all tests, while the rest complete none.
    \item ``Leaky'': $P_{\rm miss} = 1 - A$ and $P_{\rm not} = 0$, i.e. all people miss tests at random with the same probability.
\end{itemize}
These cases demonstrate how the same overall adherence to testing ($0 \leq A \leq 1$) can lead to different testing outcomes, depending on behaviour. The difference between these two behaviours can be captured by a simple model of $\Delta$IP (relative to the case with no isolation) by making the following assumptions. Suppose each individual is supposed to test every $\tau$ days and there is some window $\tau_{\rm pos}$ when they can test positive with probability $p$. If they do test positive, it will reduce their overall infectiousness by a factor $Z$. Then, for the average individual, the relative reduction in IP will be
\begin{align}
    \label{aon} \Delta \textrm{IP}_{\rm AoN} &\approx AZ\left(1-(1-p)^{\tau_{\rm pos}/\tau}\right),\\
    \label{leaky} \Delta \textrm{IP}_{\rm leaky} &\approx Z\left(1-(1-p)^{A\tau_{\rm pos}/\tau}\right).
\end{align}
Thus, $\Delta$IP is expected to scale linearly with adherence for the `all-or-nothing' case, since testing will only impact a fixed fraction of the population, while in the `leaky' case $\Delta$IP will scale non-linearly with $A$ as it will change the expected number of tests a person will take during the period when they can test positive ($A\tau_{\rm pos}/\tau$).

\subsection{Workplace contact, shift and testing patterns}
\label{sec:workplace_modelling}

To simulate the impact of testing in a workplace setting we model some simplified examples of contact and work patterns. 

\subsubsection{Modification of Infectious Potential to account for shift patterns} 
\label{sec:shift_modelling}

Equation \eqref{contact_rate} supposes that the contact rate is constant over time unless the individual is isolating. When modelling a workplace intervention, we are generally interested in the effect on workplace transmission. Therefore, to generate the results in section \ref{sec:main_results}, we consider a modified contact pattern that is proportional to work hours $c^{(w)}(t) = W(t)c(t)$ where $W(t) = 1$ during scheduled work hours and $0$ otherwise. Note, we also take $X = 1$, as we assume work contacts are completely removed by isolation. This parameterisation therefore ignores potential contacts with colleagues outside of work hours, which may also be relevant.

For simplicity, we assume that all workers do the same fortnightly shift pattern, shown in table \ref{tab:protocols}, so that all employees work, on average, 4.5 days per week, based on average working hours in the UK social care sector. 

\subsubsection{Definition of testing regimes simulated}
\label{sec:testing_regimes}

Table \ref{tab:protocols} defines the shift and testing patterns that we consider. Note that tests are assumed to take place only on work days. 
\begin{table}[ht]
    \centering
    \begin{tabular}{|l|c|c|c|c|c|c|c|}
    \hline
         Name & \multicolumn{7}{c|}{Pattern} \\ \hline
         Shift & \cellcolor{red} & \cellcolor{red} & \cellcolor{red} & \cellcolor{red} & & & \\
         pattern& \cellcolor{red} & \cellcolor{red} & \cellcolor{red} & \cellcolor{red} & \cellcolor{red} & & \\ \hline
         Daily & \cellcolor{red} & \cellcolor{red} & \cellcolor{red} & \cellcolor{red} & & & \\
         testing & \cellcolor{red} & \cellcolor{red} & \cellcolor{red} & \cellcolor{red} & \cellcolor{red} & & \\ \hline
         3 LFDs & \cellcolor{red} & & \cellcolor{red} & \cellcolor{red} & & & \\
         per week & \cellcolor{red} & & \cellcolor{red} & & \cellcolor{red} & & \\ \hline
         2 LFDs & \cellcolor{red} & & \cellcolor{red} & & & & \\
         per week & \cellcolor{red} & & & \cellcolor{red} & & & \\ \hline
         1 PCR & \cellcolor{red} & & & & & & \\
         per week & \cellcolor{red} & & & & & & \\ \hline
         1 PCR per & \cellcolor{red} & & & & & & \\
         fortnight & & & & & & & \\ \hline
    \end{tabular}
    \caption{Visual representation of the shift and testing patterns considered in sections \ref{sec:main_results}, \ref{sec:one_comp_results}, and \ref{sec:two_comp_results}. Within each row, squares from left to right indicate days of the week, upper squares indicate the first week of the pattern, and the lower squares indicate the second week. A red square means there is a shift/test scheduled for that day, while a white square indicates there is not.}
    \label{tab:protocols}
\end{table}

The numerical implementation of calculations of $\Delta$IP is detailed in \ref{app:IP_calc}. 

\subsection{Models of transmission in a closed workplace}
\label{sec:workplace_models}

To demonstrate the impact of testing interventions in a closed population, we consider two simple model workplaces. The algorithm to simulate transmission in these workplaces is detailed in \ref{app:wp_sims}.

For ease of comparison, we measure the impact of testing in these workplaces by the final outbreak size resulting from a single index case in a fully susceptible population. However in reality, mass asymptomatic testing is more useful when there is high community prevalence, and so we would expect repeated introductions over any prolonged period. We do not consider the case of repeated introductions here, nor immunity in the population, but these methods can readily be extended to that case.

\subsubsection{A single-component workplace model}
\label{sec:one_comp_model}

We consider a workplace of $N_s$ fixed staff, all of whom work the same shift pattern given by table \ref{tab:protocols}. Each individual's shift pattern starts on a random day (from 1 to 14) so it is assumed that approximately the same number of workers are working each day. Contacts for each infectious individual on shift each day are drawn at random from the rest of the population on shift that day with fixed probability $p_c$. Each contact is assumed to have probability of infection
\begin{align}
    \label{p_inf} p_{k,k'}(t) = 1 - \exp\left[-\beta_0 J_k(t) s_{k'}(t)\right]
\end{align}
where $\beta_0$ is the (average) transmission rate for the contact, $J_k(t)$ is the infectiousness of the infectious individual ($k \in \{1,\ldots,N_s\}$), $s_{k} \in \{0,1\}$ is the susceptibility of the contact ($k' \in \{1,\ldots,N_s\} \neq k$).

An upper bound for the approximate reproduction number in this workplace can be calculated as follows
\begin{align}
    R_{\rm wp} \lesssim f_s^2 \beta_0 p_c (N_s-1) \langle \tau_{\rm inf}\rangle (1-\Delta\textrm{IP})
\end{align}
where $f_s$ is the fraction of days on-shift (9/14 here) and $1 - \Delta$IP is the relative infectious potential after taking into account any testing and isolation measures. 

\subsubsection{A two-component model to represent transmission in a care-home}
\label{sec:two_comp_model}

We also extend the model in the previous section to a basic model of contacts in a care home consisting of two populations: staff and residents. We assume the same shift patterns apply for the staff as in the previous section, but that residents are present in the care home on all days. As in the previous section, we assume that contacts are drawn at random each day for infectious individuals from the pool of other individuals at work that day. However, the contact probability is different for resident-resident, staff-resident, and staff-staff contacts such that it can be expressed by the following matrix
\begin{align}
    \textbf{P}_c = p_c \begin{pmatrix}
    a & 1 \\
    1 & b
    \end{pmatrix},
\end{align}
such that $p_c$, $a p_c$, and $b p_c$ are the staff-resident,  staff-staff, and resident-resident contact probabilities respectively. We assume that transmission dynamics are the same for all groups but that testing and isolation measures can be applied separately. It is important to note that here we only consider the case of a closed population with a single (staff) index case, which is most similar to the early pandemic (i.e. a fully susceptible population, low incidence, and staff ingress is more likely than patient ingress due to limits on visitors). The situation becomes complex in more realistic scenarios (e.g. \cite{rosello_2022}) however many of the lessons we can learn from this simple case are transferable (at least qualitatively).

In section \ref{sec:two_comp_results} we consider two cases to measure the impact of mass asymptomatic testing of staff. First, to mimic the impact of social distancing policies for staff we vary $a$ while keeping $p_c$ and $b$ constant (i.e. minimising disruption for residents) and compare the absolute and relative impacts of testing interventions. Second, to show the effect of the relative contact rates, which may vary widely between individual care-homes, we vary $a$ and $b$ such that the average number of contacts an infectious person will make is fixed, meaning that we impose the following constraint
\begin{align}
    \label{res_contact} b = 1 - \frac{N_sf_s(N_sf_s - 1)}{N_r(N_r - 1)}(a-1)
\end{align}
where $N_r$ is the size of the fixed resident population, and $N_s$ is the size of the fixed staff population. We will use this to investigate how testing policies can have different impacts in different care-homes even if the have similar overall transmission rates or numbers of cases. 

\section{Results}

\subsection{Role of population heterogeneity in testing efficacy}
\label{sec:correlation}

The individual viral-load based model introduced here accounts for correlations between infectiousness and the probability of testing positive both between individuals and over time. For example, people with a higher peak viral load are more likely to be positive and more likely to be (more) infectious. In this section we demonstrate how the model assumptions around heterogeneity in, and correlations between, peak viral load and peak infectiousness affect predictions of testing efficacy. To do this, we first calculate a population average model, which uses the time-point average of the population of $N$ individuals for the infectiousness and test-positive probability
\begin{align}
    \langle J(t)\rangle &= \frac{1}{N}\sum_{k=1}^{N} J(V_k(t)), \\
    \langle P_{\rm LFD}(t)\rangle &= \frac{1}{N}\sum_{k=1}^{N} P_{\rm LFD}(V_k(t)).
\end{align}
These profiles are then assigned to all individuals in a parallel population of $N$ individuals so that the impact of population heterogeneity can be compared directly.

\begin{figure}[ht]
\centering
\includegraphics[width=\textwidth]{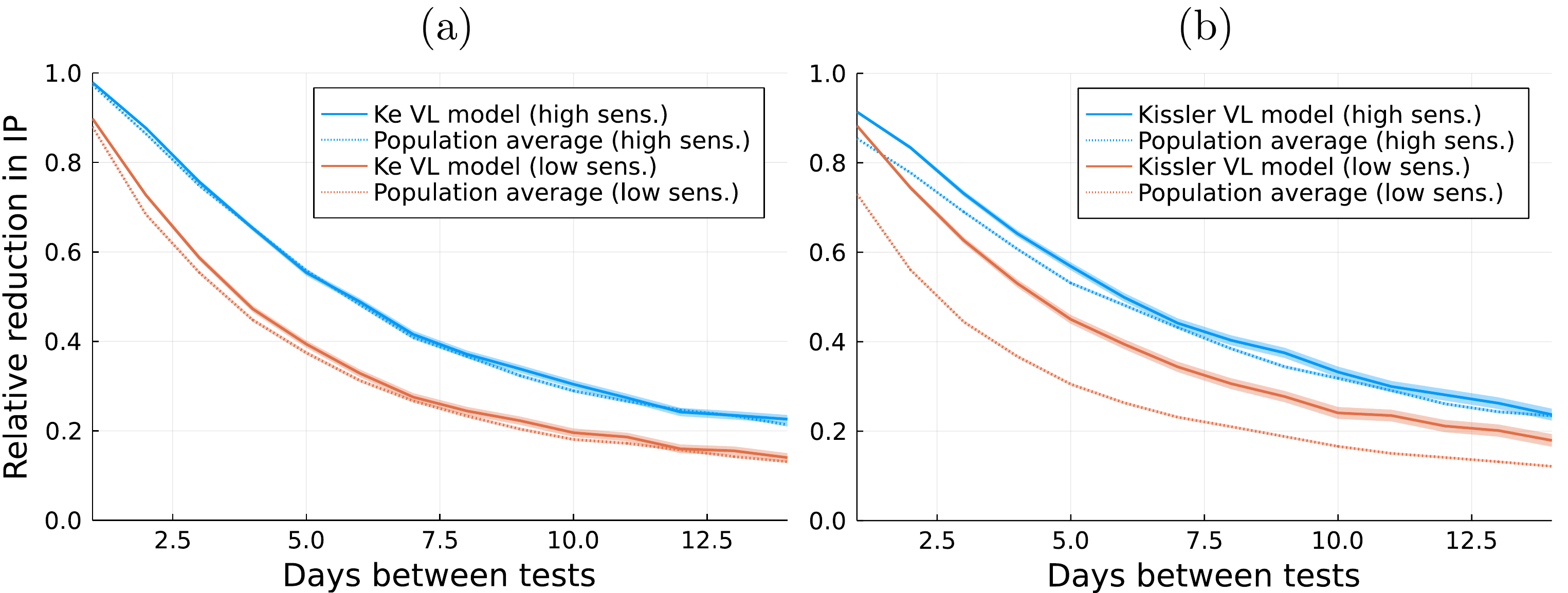}
\label{fig:correlation}
\caption{Plots of $\Delta$IP, the relative change in IP due to regular asymptomatic testing with LFDs vs. the time between tests. (a) Results using the Ke et al. based model of RNA viral load and (b) using the Kissler et al. based model. Blue lines show the results using the `high' sensitivity model for LFD testing and orange the `low sensitivity model'. Dashed lines show the results using a population average model, based on using the population mean infectiousness and test-positive probability for all individuals. In all cases the reduction is calculated as in equation \eqref{DeltaIP} relative to the baseline case with no testing and no isolation at symptom onset ($P_{\rm isol} = 0$). Total adherence to all testing regimes is assumed in this case, and each point is calculated using the same population of 10,000 generated individuals. The shaded areas show the 95\% confidence intervals in the mean of $\Delta$IP, approximated by 1000 bootstrapping samples.}
\end{figure}

Figure \ref{fig:correlation} compares the relative overall change in infectious potential ($\Delta$IP) for a population of infected individuals performing LFDs at varying frequency for the heterogeneous models vs their homogeneous (population average) counterparts. This is shown for several cases; in figure \ref{fig:correlation}(a) the Ke et al. based model is used, which has low population heterogeneity in the PE parameters. Therefore, we see little difference between the full model prediction and the population average case. The `high' sensitivity testing model outperforms the `low' sensitivity model, as expected. However, the difference is proportionally smaller for high testing frequencies because frequent testing can compensate for low sensitivity to some extent. Furthermore, in the `high' sensitivity model subjects are likely to test positive for longer periods of time and so retain efficacy better at longer periods between tests. 

Comparing this to figure \ref{fig:correlation}(b) we see that there is a much larger discrepancy between the Kissler et al. model and its population average. This is because there is much more significant population heterogeneity in this model, and the correlation between infectiousness and test positive probability means that individuals who are more infectious are more likely to test positive and isolate. Therefore the effect of testing is significantly larger than is predicted by the population average model. This effect is also much larger in the `low' sensitivity model, because in this model the test-positive probability has a very similar RNA viral load dependence to the infectiousness (with approximately $10^6$ copies/ml being the threshold between low/high test-positive probability and infectiousness). This means there is greater correlation between these two properties and so this effect is amplified.

The between-individual relationship of infectiousness and viral load for SARS-CoV-2 is still largely unknown. While studies have shown a correlation between viral load and secondary cases \cite{marks_transmission_2021,lee_2022}, this could also be affected by the timing of tests (which may also be correlated to symptom onset) and therefore the within-individual variability in viral load over time. We found in this section that even though the two models of RNA viral load we use predict very similar testing efficacy, they highlight important factors to consider when modelling repeat asymptomatic testing:
\begin{enumerate}
    \item Population heterogeneity: greater heterogeneity means poorer agreement between the individual-level and population-average model predictions.
    \item Between individual correlation of infectiousness and test-positive probability: the greater this correlation is the more important the heterogeneity is to predictions of test efficacy.
\end{enumerate}
Thus, quantifying population heterogeneity in infectiousness (i.e. super-spreading) and likelihood of testing positive while infectious and the correlation between the two can significantly affect predictions of efficacy of repeat asymptomatic testing.

\subsection{Modelling the impact of non-adherence}
\label{sec:adherence_results}

\begin{figure}[ht]
\centering
\includegraphics[width=\textwidth]{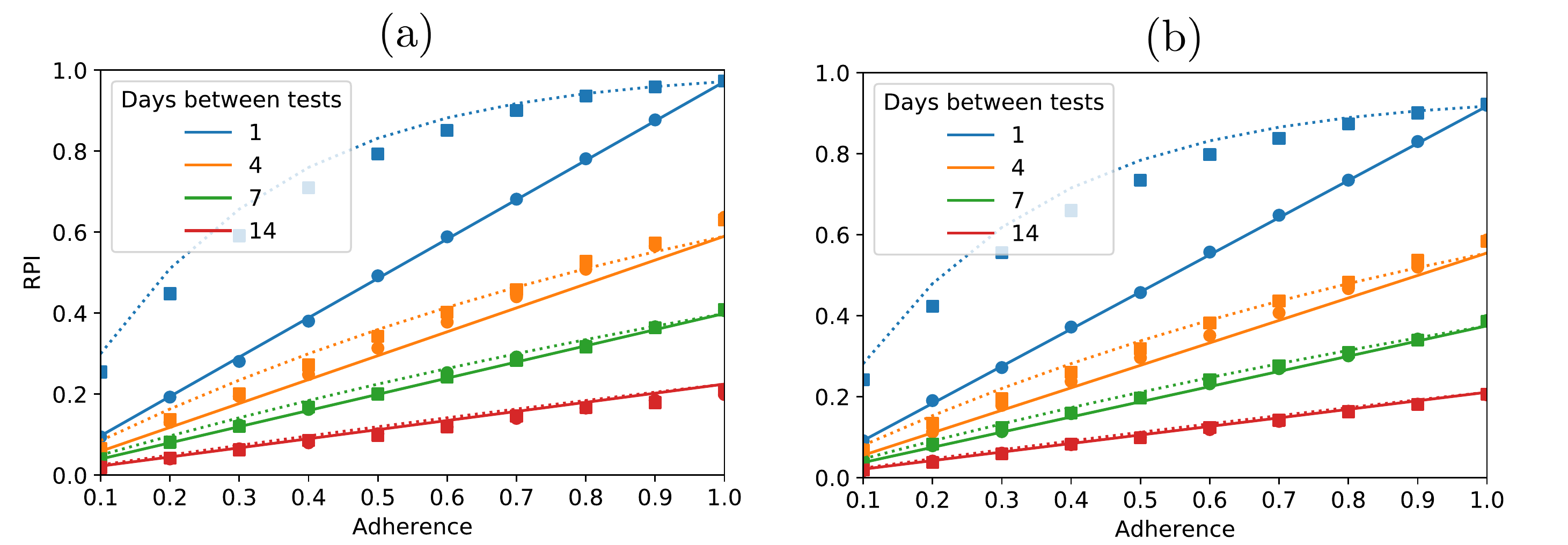}
\caption{Relative reduction in IP ($\Delta$IP) vs. adherence calculated for the (a) Ke et al. based model of RNA viral load and (b) Kissler et al. based model. The circles show the results when adherence is `all-or-nothing' while squares show the case when it is `leaky'. Error bars show the 95\% confidence intervals of the mean approximated by 1000 bootstrapping samples. Additionally, the solid lines show equation \eqref{aon} while the dashed lines show equation \eqref{leaky} with parameters given by the final column of table \ref{tab:adherence}. The dot and line colours correspond to different testing frequencies, as labelled in the captions. In all cases the reduction was calculated as in equation \eqref{DeltaIP} relative to the baseline case with no testing and no isolation at symptom onset ($P_{\rm isol} = 0$). Each point was calculated using the same population of 10,000 generated individuals and the `high' sensitivity model of LFDs was used. }
\label{fig:adherence}
\end{figure}

In figure \ref{fig:adherence} we calculate the two extremes of adherence behaviour considered here, comparing the ``all-or-nothing'' and ``leaky'' adherence models. We see that these different ways of achieving the same overall adherence only differ noticeably when testing at high frequency (as shown by the results for daily testing in figure \ref{fig:adherence}). At high frequency, `leaky' adherence results in a greater reduction in IP, because even though tests are being missed at the same rate, all individuals are still testing at a high-rate and so have a high chance of recording a positive test. Conversely, in the `all-or-nothing' case, obstinate non-testers can never isolate, so the changing test frequency can only impact that sub-population who do test. 

We also see from figure \ref{fig:adherence} that the relative reduction in IP ($\Delta$IP) is well approximated by fitting the reduced model of testing in equations \eqref{aon} and \eqref{leaky} to the data for $\Delta$IP. The fitted parameters for the two viral load models are given in table. We fit the models for the `all-or-nothing' and `leaky' cases separately to the $\Delta$IP data using a least-squares method. Table \ref{tab:adherence} shows that the two cases give very similar testing parameters, suggesting that the simple model captures the behaviour well. The solid lines in figure \ref{fig:adherence} show the simple model results (equations \eqref{aon} and \eqref{leaky}) using the mean fitted parameters from the final column of table \ref{tab:adherence}.
\begin{table}[ht]
    \centering
    \begin{tabular}{|l|l|c|c|c|c|c|c|}
        \hline RNA viral & Adherence & \multicolumn{3}{c|}{Fitted parameters} & \multicolumn{3}{c|}{Mean of fitted parameters} \\ \cline{3-8} 
        load model & behaviour & Z & p & $\tau_{\rm pos}$ (days) & Z & p & $\tau_{\rm pos}$ (days) \\ \hline
        \multirow{2}{*}{Ke et al.} & `All-or-nothing' & 1.0 & 0.562 & 4.47 & \multirow{2}{*}{1.0} & \multirow{2}{*}{0.558} & \multirow{2}{*}{4.37} \\ \cline{2-5} 
        & `Leaky' & 1.0 & 0.555 & 4.26 & & & \\ \hline
        \multirow{2}{*}{Kissler et al.} & `All-or-nothing' & 0.958 & 0.561 & 4.38 & \multirow{2}{*}{0.945} & \multirow{2}{*}{0.562} & \multirow{2}{*}{4.28} \\ \cline{2-5} 
        & `Leaky' & 0.932 & 0.563 & 4.19 & & &\\ \hline
    \end{tabular}
    \caption{Parameters of the simplified models of $\Delta$IP given in equation \eqref{aon} and \eqref{leaky} fitted to the scatter plot data in figure \ref{fig:adherence}. The parameters were fitted separately for each viral load model and each model of adherence behaviour. The final column shows the mean parameters from the `all-or-nothing' and `leaky' model fits, which were used to generate the line data in figure \ref{fig:adherence}.}
    \label{tab:adherence}
\end{table}

To summarise, a single adherence parameter may be sufficient to capture how adherence affects the impact that regular testing has on infectious potential, but only when the testing is not very frequent (e.g. every 3 or more days). When testing is frequent, the very simple model of equations \eqref{aon} and \eqref{leaky} can be used to estimate the potential impact of testing at different frequencies on infectious potential for two extreme models of behaviour. Namely, when adherence is `leaky' or `all-or-nothing'. 

\subsection{Comparison of staff testing policies for high-risk settings}
\label{sec:main_results}

In the previous sections we have considered simple testing strategies consisting of repeated LFDs at a fixed frequency. In this section we consider scenarios more relevant to workplaces, outlined in table \ref{tab:protocols}. Due to the mix of test types, it is less clear \emph{a priori} how the regimes will compare in efficacy. A key question we consider is whether substituting a weekly PCR test with an extra LFD test results in the better, worse, or similar $\Delta$ IP, depending on the underlying assumptions.

\begin{figure}[ht]
\centering
\includegraphics[width=\textwidth]{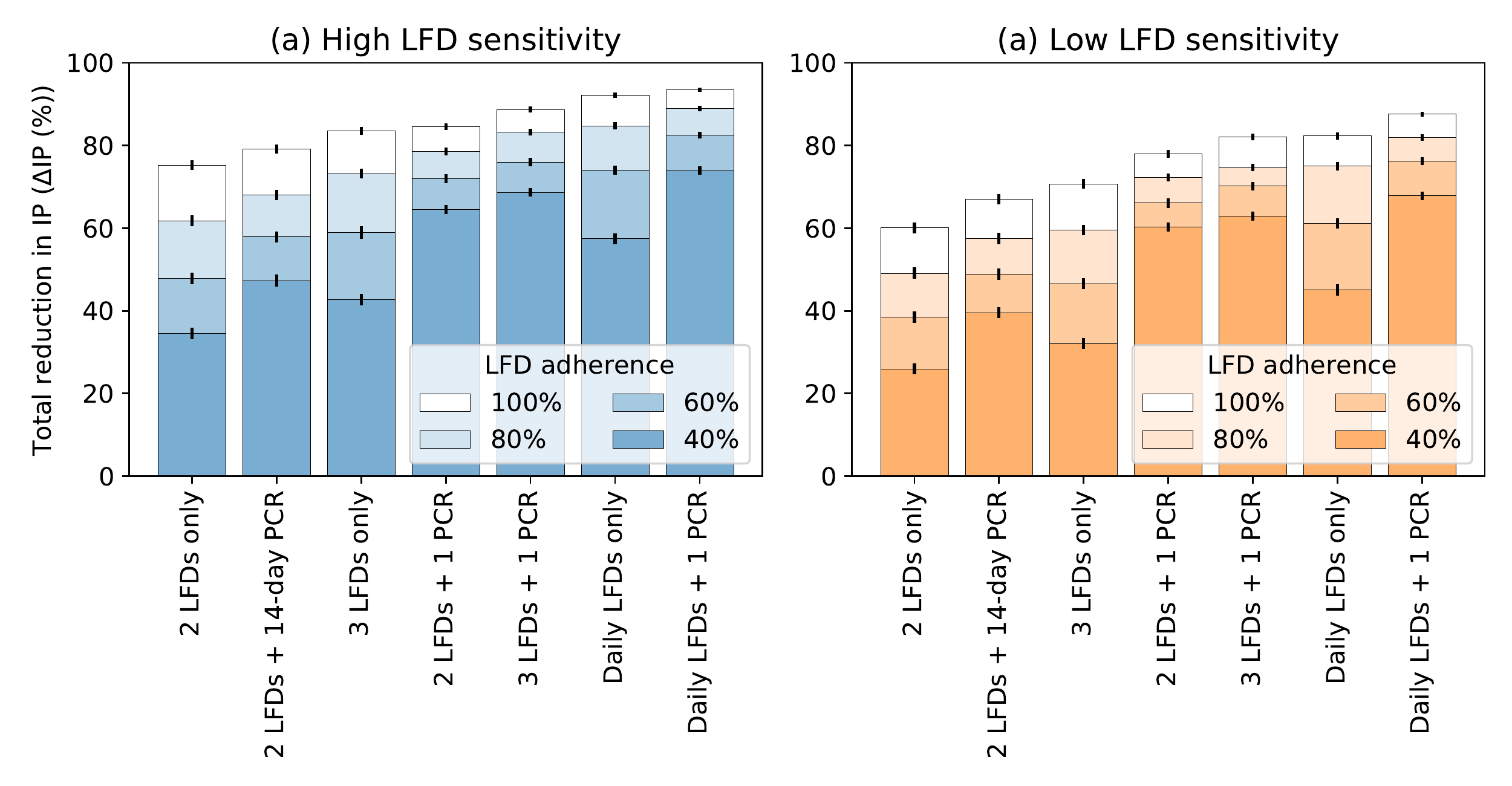}
\caption{Reduction in population infectious potential expressed as a percentage relative to a baseline case with no testing and symptom isolation with probability $P_{\rm isol} = 1.0$. Testing regimes simulated are from left to right in order of their effectiveness at 100\% adherence. (a) and (b) only differ in the model of LFD sensitivity used, as labelled. Each bar is calculated using 10,000 samples, lighter coloured bars are used to show the extra $\Delta$IP gained by increasing LFD adherence from a baseline of 40\%. A mean PCR turnaround time of 45h is assumed. Error bars indicate 95\% confidence intervals of the mean, approximated using 1000 bootstrapping samples.}
\label{fig:scenarios}
\end{figure}

\begin{figure}[ht]
\centering
\includegraphics[width=\textwidth]{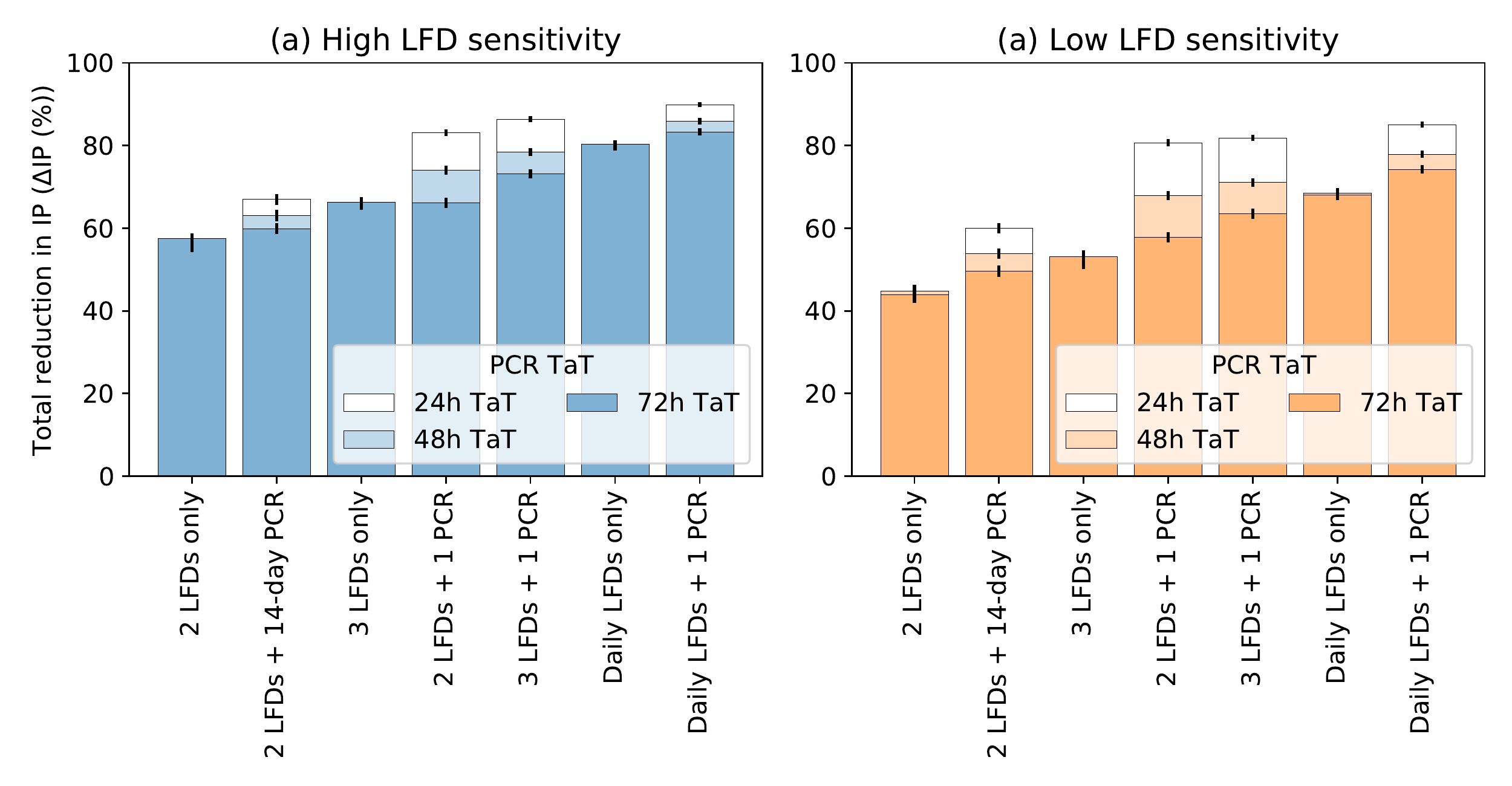}
\caption{The same plot as figure \ref{fig:scenarios} except the lighter bars show the effect of reducing PCR turnaround time (TaT) from a baseline of 72h. A `leaky' adherence of 70\% is assumed. Error bars indicate 95\% confidence intervals of the mean, approximated using 1000 bootstrapping samples.}
\label{fig:tat}
\end{figure}

Figure \ref{fig:scenarios} shows the main results comparing the various regimes. At 100\% adherence and high LFD sensitivity (figure \ref{fig:scenarios}(a)), we find some interesting results, primarily that ``2 LFDs + 1 PCR'' and ``3 LFDs'' perform similarly, as do ``3 LFDs + 1 PCR'' and ``Daily LFDs'', suggesting that, in theory, substituting PCR tests for LFD tests does not have a large impact on transmission reduction. This is because, even though PCR tests are more sensitive than LFDs, the turnaround time from taking a PCR test to receiving the result limits the potential reduction in IP that can be achieved by these tests. This is demonstrated in figure \ref{fig:tat} by the change in $\Delta$IP as we change the mean PCR turnaround time. In the low sensitivity case (figure \ref{fig:scenarios}(b)), there is a larger difference between ``2 LFDs + 1 PCR'' and ``3 LFDs'', but again ``Daily LFDs'' outperform  ``3 LFDs + 1 PCR'' (at 100\% adherence) since the high-frequency of testing counteracts the low sensitivity by providing multiple chances to test positive. 

Another important implication of figure \ref{fig:scenarios} is the effect of varying LFD adherence (in this case assuming `leaky' adherence behaviour). Naturally, this impacts much more strongly on the LFD-only regimes demonstrating the usefulness of the PCR tests as a less-frequent but mandatory and highly sensitive test as a buffer in case LFD adherence is low or falling. Another factor to consider when changing or between testing regimes is how this will affect adherence levels. For example, if the workforce is performing 60\% of the LFD tests that are set out by the testing regime, but then the regime is changed from `2 LFDs + 1 PCR' to `Daily LFDs only', the adherence rates are likely to fall. In this case, the results in figure \ref{fig:scenarios} can be used to estimate how much they would need to fall for $\Delta$IP to go down (for this example, only in the region of approx 5-10\%, assuming high LFD sensitivity and `leaky' adherence). Note that, if `all-or-nothing' adherence was used instead, the impact of non-adherence for the `LFD only' regimes would be even greater than shown in figure \ref{fig:scenarios}, due to the arguments outlined in section \ref{sec:adherence_results}.

\begin{figure}[ht]
\centering
\includegraphics[width=\textwidth]{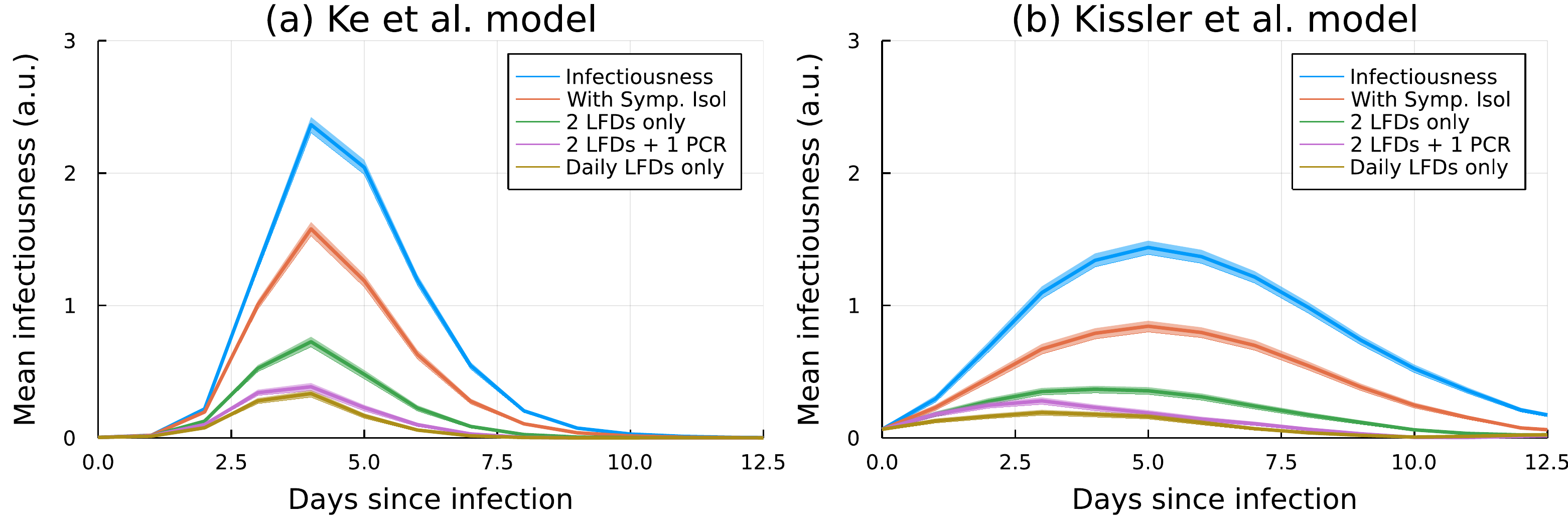}
\caption{Population mean values $\langle J_t \rangle$ under different testing and isolation regimes. The blue line shows the baseline infectiousness, with no isolation, and the orange line shows the case with symptomatic isolation with perfect adherence $P_{\rm isol} = 1.0$. The testing regimes simulated are labelled in the caption and are all simulated assuming `leaky' adherence at 70\%. (a) Shows the results using the Ke et al. model of RNA viral load and (b) the Kissler et al. model. All curves are calculated using 10,000 samples and the 95\% confidence intervals on the mean is given by the shaded area (when this is thicker than the line). }
\label{fig:IPvis}
\end{figure}

Finally, figure \ref{fig:IPvis} shows what happens to the population average infectiousness under different testing regimes. As testing frequency is increased, the bulk of infectiousness is pushed earlier in the infection period, as individuals are much more likely to be isolated later in the period. This is an example of how testing and isolation interventions can not only reduce the reproduction number, but also the generation time, as any infections that do occur are more likely to occur earlier in the infectious period.

To summarise, we find that the effect of LFD and PCR tests are comparable when PCR tests have a $\sim 2$-day turnaround time, in line with other studies \cite{hellewell_2021,quilty_2021,larremore_test_2021}. However, differential adherence is likely to be the key determinant of efficacy when switching a PCR for LFD. Observed rates of adherence to workplace testing programmes will likely depend on numerous factors including how the programme is implemented, the measures in place to support self-isolation and the broader epidemiological context (i.e. prevalence and awareness). 

There is uncertainty in the parameters used to make these predictions, so to quantify the impacts of parameters uncertainty on $\Delta$IP by performing a sensitivity analysis, which is presented in \ref{app:sens_analysis}. This shows that certain parameters are less important, such as the coupled timing on peak viral load and symptom onset, or the viral load growth rate. Unsurprisingly, the $\Delta IP$ predictions for 2 LFDs per week is most sensitive is most sensitive to LFD sensitivity parameters $\lambda$ and $\mu_l$. However, the daily LFDs case is most sensitive to the infectiousness parameter $h$. This is because the sensitivity of individual (independent) tests becomes less important as they are repeated regularly, and a key determinant of IP then is the proportion of infectiousness that occurs in the early stages of infection, before isolation can feasibly be triggered, which increases with smaller $h$. Interestingly however, the infectiousness threshold parameter $K_m$ does not seem to have as large an effect. In part this is because it has less uncertainty associated with it, but $h$ also has a more profound effect is because it not only does decreasing it increase pre-symptomatic infectiousness (as does decreasing $K_m$), it also reduces the relative infectiousness around peak viral load, thereby decreasing the value of LFD-triggered isolation (which is mostly likely to start near to peak viral load).

PCR tests for SARS-CoV-2 generally come at a much higher financial cost than LFD tests because of the associated lab costs, and so are not feasible for sustained deployment by employers or governments. In this context, the potential impact of regular LFD testing is clear and sizeable, so long as policy adherence can be maintained. The two models of LFD test sensitivity change the results, but qualitatively we see that if all people perform 2 LFDs per week (either `2 LFDs' regime at 100\% adherence or '3 LFDs' at $\sim 70\%$ adherence), then the reproduction number can be halved (at least) and potentially reduced by up to 60-70\%. This is a sizeable effect and so regular asymptomatic with LFDs is potentially a cost-effective option at reducing transmission in workplaces.


\subsection{Infectious Potential as a predictor of transmission in a simple workplace model}
\label{sec:one_comp_results}

\begin{figure}[ht]
\centering
\includegraphics[width=0.9\textwidth]{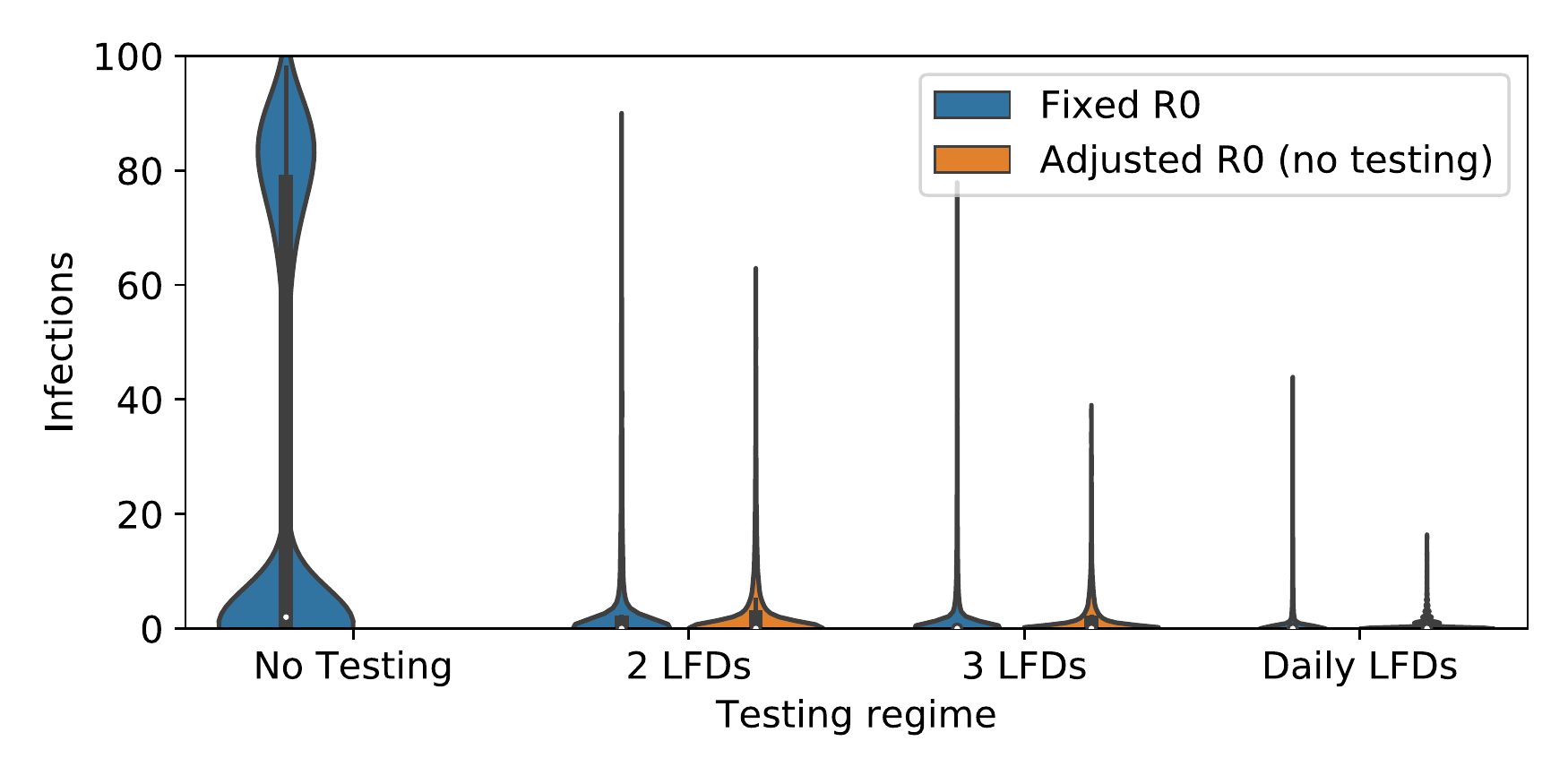}
\caption{Violin plots showing the distribution of the outbreak size in a workplace of $N=100$ people given a single index case. Blue violins show the cases where testing is modelled explicitly (except in the `no testing' case), while orange violins show cases with no testing but where the contact rate $p_c$ is reduced by a factor $\Delta$IP to mimic the testing regime in question. Each violin consists of 10,000 simulations and the white dot shows the median of the distribution. A `leaky' adherence to testing at 70\% was assumed. The case shown uses the Ke et al. model of RNA viral load and the `high' sensitivity model of LFD testing. The transmission parameters used were $p_c=0.296$ and $\beta_0=0.0265$ giving an approximate $R_{wp}$ value of 3 (with symptom isolation). Note that the `no testing' case still includes symptomatic isolation with $P_{\rm isol} = 1.0$, and so the baseline $R_0$ is not realised. }
\label{fig:one_comp}
\end{figure}

It is shown in equation \eqref{Rnum} how IP is related to the reproduction number under some simplifying assumptions about transmission. Figure \ref{fig:one_comp} demonstrates that this relationship is well approximated even in the stochastic model outline in section \ref{sec:one_comp_model}. It compares the probability distributions of outbreak sizes (resulting from a single index case) in a closed workplace of 100 employees under the different LFD only testing regimes. These are presented next to the same results for a model with no testing, but with a reduced contact rate $p_c \rightarrow (1 - \Delta$IP$)p_c$ where $\Delta$IP is the reduction in IP predicted for the corresponding testing regime. In other words, the baseline $R_0$ value of the workplace is adjusted to match what would be expected if a particular testing regime was in place.

We see that the final outbreak sizes are fairly well predicted, even though temporal information about infectiousness (shown in figure \ref{fig:IPvis}) is not captured by the simpler model. The key difference between explicit simulations of the testing regimes (in blue) and the approximated versions (in orange) is the heterogeneity in outcomes. Testing is a random process and leads to greater heterogeneity in infectious potential, by simply scaling transmission by the population level $\Delta$IP, that heterogeneity is lost. 

Nonetheless, these results demonstrate the usefulness of $\Delta$IP as a measure. All of the testing regimes simulated in figure \ref{fig:one_comp} reduce the workplace reproduction number to less than the critical value for this stochastic model with $N=100$ employees, and this is matched by the predictions given by $\Delta$IP. Therefore, given some data or model regarding the baseline transmission rate in the setting of interest, calculating $\Delta$IP is an efficient way of approximating the impact of potential testing interventions and predicting how frequent testing will have to be to reduce the reproduction number below a threshold value (generally $\gtrsim 1$ for finite-population models \cite{ball_1994}).

\begin{figure}[ht]
\centering
\includegraphics[width=\textwidth]{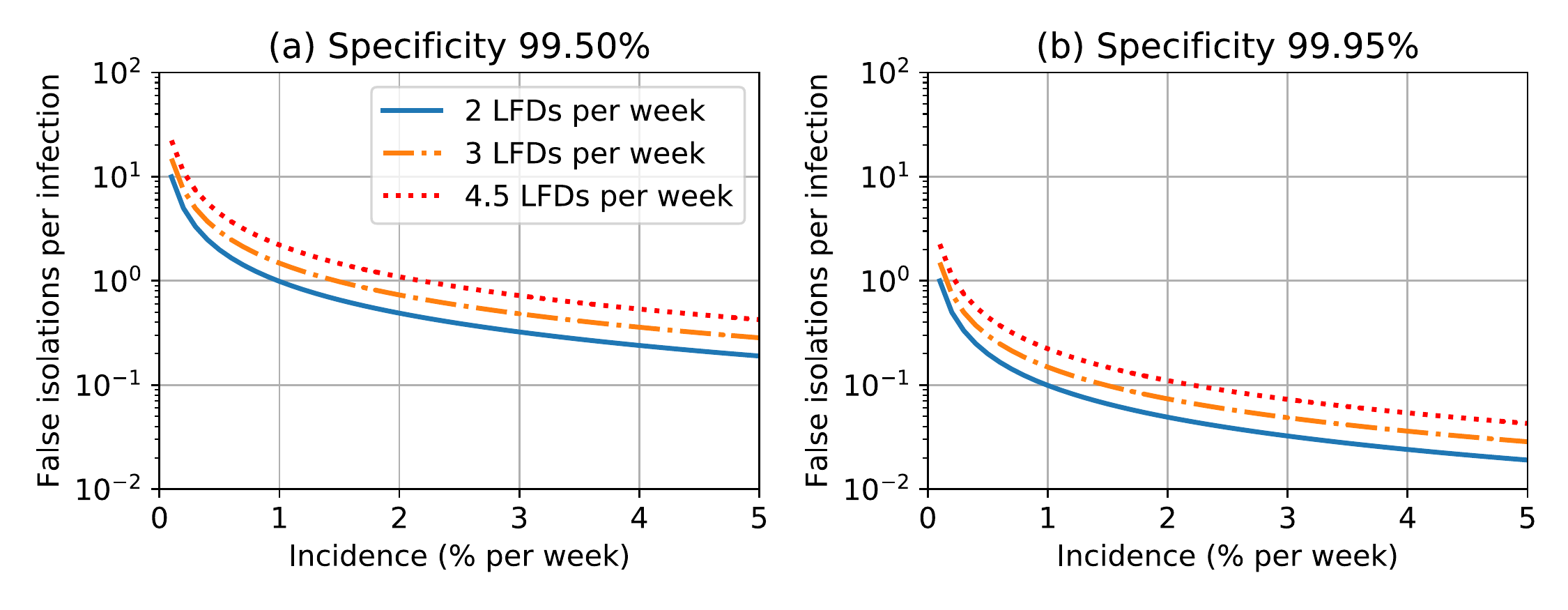}
\caption{Average number of false positives per new infection in the population at different rates of incidence. (a) 99.5\% specificity, similar to that reported in \cite{peto_2021}. (b) 99.95\% specificity, similar to that reported in \cite{wolf_2021}.}
\label{fig:false-pos}
\end{figure}

False-positives are also an important factor when considering the costs of any repeat testing policies as even tests with relatively high specificity performed frequently enough will produce false-positive results. Figure \ref{fig:false-pos} shows a direct calculation of the number of false positives per actual new infection in the population for the testing regimes considered here. The figure demonstrates how a small change in specificity greatly changes the picture. The case in figure \ref{fig:false-pos}(b) is close to more recent estimates of LFD specificity \cite{wolf_2021}, suggesting that the rate of false-positives will only become comparable to the number of new infections at very low incidence. Imposing some threshold on this quantity is a measure of how many false positives (and impacts thereof) the policy-maker is willing to accept in search of each infected person.



\subsection{Testing to protect vulnerable groups in a two-component work-setting}
\label{sec:two_comp_results}

The simple picture of IP$\propto R_{\rm wp}$ becomes less straight-forward as we consider workplaces of increasing complexity. In this section we consider the reduced model of a care-home outlined in section \ref{sec:two_comp_model}. We model the case where the index case is a staff member (as generally residents are more isolated from the wider community \cite{rosello_2022}) and testing policy only applies to staff.

\begin{figure}[ht]
\centering
\includegraphics[width=\textwidth]{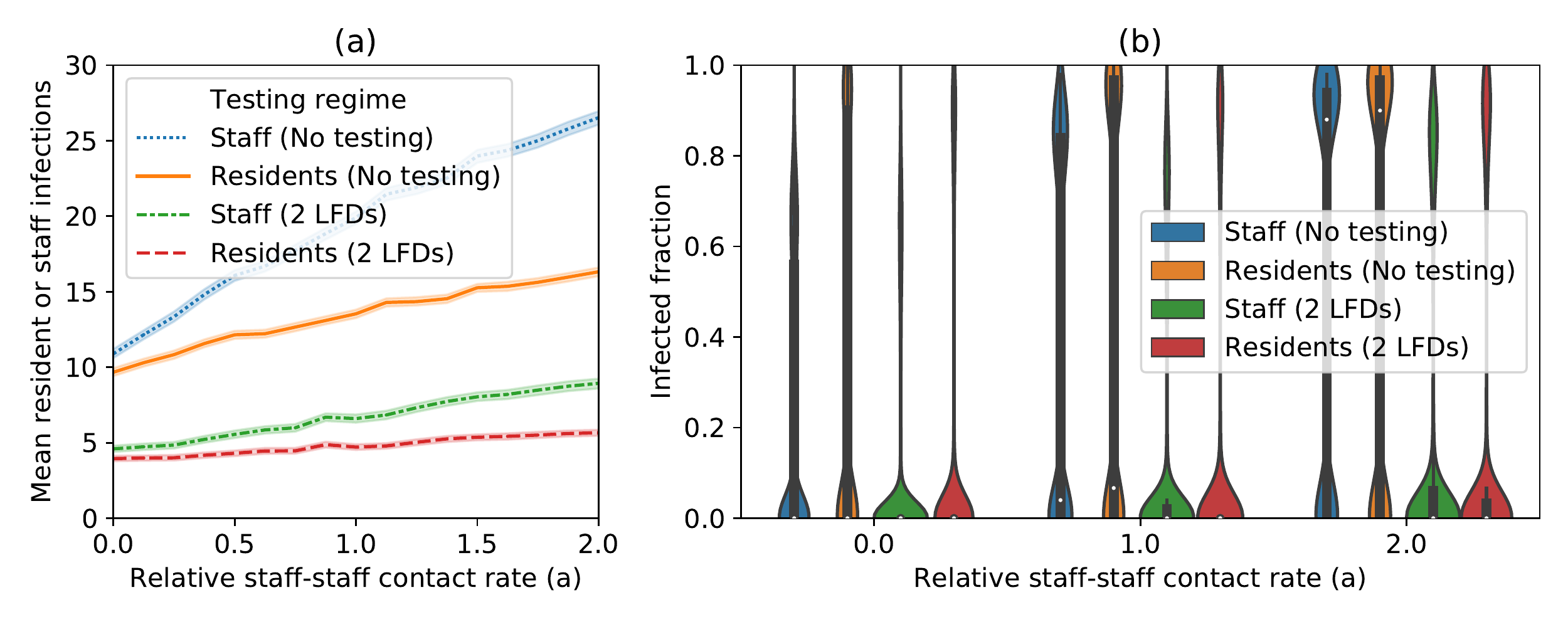}
\caption{Summary of staff and resident infections in the two-component model of care-home contacts while varying the relative staff-staff contact rate $a$ and fixing the relative resident-resident contact rate $b=1$. The index case for the outbreak was assumed to be a staff member. (a) The mean number of residents and staff infected in simulations given the staff-staff contact rate $a$, with and without staff testing of 2 LFDs per week (as labelled). The shaded area indicates the 95\% confidence intervals in the mean. (b) Violin plots of the resident and staff infections in the same scenarios, divided by the total number of residents and staff respectively, for select values of $a$. The parameters used to generate these plots were: total number of residents $N_r = 30$  and staff $N_s = 50$, contact probability $p_c=0.296$, and transmission rate $\beta_0=0.0265$. Also, the Ke et al. model of RNA viral load and the `high' sensitivity model of LFD testing. 10,000 simulations were realised to generate these results and a `leaky' adherence to testing at 70\% was assumed.}
\label{fig:two_comp_avary}
\end{figure}

Figure \ref{fig:two_comp_avary} shows the effect of varying the staff-staff contact rate $a$. As expected, reducing $a$ reduces the reproduction number and therefore the final outbreak size, demonstrating how social distancing of staff alone would reduce both staff and resident infections, but have a larger effect on staff infections (figure \ref{fig:two_comp_avary}(a)). Regular asymptomatic staff testing is predicted to have a sizeable effect on resident infections, reducing them by 50-60\% across the whole range of $a$. Interestingly, in the presence of this effective staff testing intervention, staff social distancing is predicted to have a minimal effect on resident infections. When $a$ is very small (i.e. staff don't interact) most transmission chains will have to involve residents to be successful, hence we see similar infection rates for both groups. At large $a$, staff outbreaks become more common than resident outbreaks (figure \ref{fig:two_comp_avary}(b)), however by both reducing staff-staff transmission, and screening residents from infectious staff, staff testing has a larger relative effect on infections in both groups.

\begin{figure}[ht]
\centering
\includegraphics[width=\textwidth]{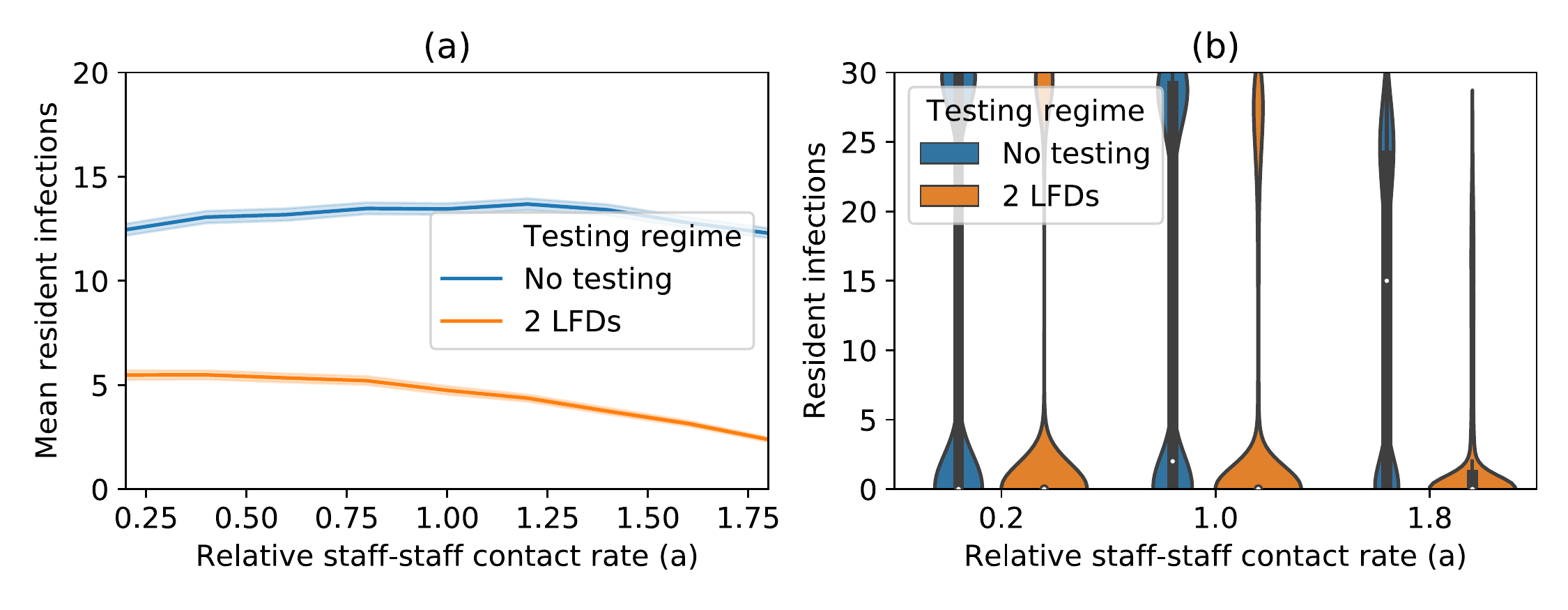}
\caption{Summary of resident infections in the two-component model of care-home contacts where the staff-staff and resident-resident contact rates ($a$ and $b$ respectively) are varied simultaneously as shown in equation \eqref{res_contact}. The index case for the outbreak was assumed to be a staff member. (a) The mean number of residents infected in simulations given the staff-staff contact rate $a$, with (blue) and without (orange) staff testing of 2 LFDs per week. The shaded area indicates the 95\% confidence intervals in the mean. (b) Violin plots of the resident infections in the same scenarios, for select values of $a$. All parameters other than $a$ and $b$ are the same as in figure \ref{fig:two_comp_avary}.}
\label{fig:two_comp}
\end{figure}

In figure \ref{fig:two_comp} we focus on the dependence of resident infections on the underlying contact structure, by varying $a$ and $b$ simultaneously to maintain a constant reproduction number in the care-home (see equation \eqref{res_contact}). Interestingly, under these constraints, when staff undergo regular testing (in this case the `2 LFDs' regime), there is a much stronger (negative) dependence of resident infections on $a$ observed. This is because infectious staff are more likely to isolate due to the testing and so then resident-resident contacts become the key route for resident infections. Therefore, with staff testing, higher resident-resident contact rates (decreasing $a$) increases resident cases because outbreaks can occur in this population relatively unchecked. 

Therefore, while $\Delta$IP is a useful measure for comparing different testing regimes, in the more complex setting of a care home an understanding of the underlying transmission rates between and within staff and resident populations is required to understand how this affects the probability of an outbreak. Nonetheless, given knowledge of these underlying contact/transmission rates, $\Delta$IP can still be very useful as a measure of efficacy. In the case presented here, only staff undergo testing and so only staff$\rightarrow$staff and staff$\rightarrow$resident transmission are reduced by a factor of approximately $\Delta$IP. This means that while staff testing will reduce the number of resident infections resulting from a new staff introduction of SARS-CoV-2 into the workplace, the relationship between $\Delta$IP and resident infections is more complex, limiting the scope of policy advice that can be given for this setting based on $\Delta$IP alone.

\section{Discussion}

This paper presents a simple viral load-based model of the impact of asymptomatic testing on transmission of SARS-CoV-2, particularly for workplace settings, by using data from repeat dual-testing data in the literature. The results here highlight several important aspects for both modelling testing interventions and making policy decisions regarding such interventions. 

In terms of modelling implications, in section \ref{sec:correlation}, we highlighted that a combination of population heterogeneity and correlation between test-positive probability and infectiousness will increase the overall predicted effect of testing interventions. In short, this is because if people who are more infectious are also more likely to test positive then testing interventions become more efficient at reducing transmission. In section \ref{sec:adherence_results} we also showed that model predictions can be affected by assumptions around adherence behaviour. In an analogy to models of vaccine effectiveness, we consider two extremes of adherence behaviour, ``all-or-nothing'' and ``leaky'' adherence. Testing is always more effective in a population with leaky adherence (assuming the same overall adherence rate) but the difference between the two cases is only predicted to be significant when testing very frequently (every 1-2 days). Real behaviour is more nuanced than these extremes, and in a population at any one time will likely consist of a continuum of rates of adherence. Nonetheless, highlighting these extremes is important for giving realistic uncertainty bounds for cases when only an overall adherence rate is reported, and for understanding the impact of assumptions that are implicit in models of testing.

As for policy implications, in section \ref{sec:main_results} we demonstrate that regular testing can be highly effective at reducing transmission assuming that adherence rates are high. This work suggests that regular testing with good adherence could control outbreaks in workplaces with a baseline $R_{wp} \sim 3$ (sections \ref{sec:one_comp_results} and \ref{sec:two_comp_results}). Estimates of the basic reproduction number for SARS-CoV-2 are in the range 2-4 for the original strain \cite{du_2022} and up to $10$ for Omicron variants \cite{liu_2022}. Of course the effective reproduction number in specific work-settings may is likely to be lower, depending on the frequency and duration of contacts and symptom isolation behaviour. 

This paper also highlights that the level of adherence with testing interventions is crucial to their success and also one of the most difficult factors to predict in advance. Numerous factors determine how people engage with testing and self-isolation policies including the cost of isolation (e.g. direct loss of earnings) \cite{smith_2021} and perceived social costs of a positive test (e.g. testing positive may require co-habitants to isolate too) \cite{michie_2020,blake_2021}. In studies of mass asymptomatic testing of care-home staff it was found that increasing testing frequency reduced adherence \cite{tulloch_2021}, and also added to the burden of stress felt by a workforce already overstretched by the pandemic \cite{kierkegaard_implementing_2021}. Therefore the results of modelling studies such as this paper need to be considered in the wider context of their application by decision makers, and balanced against all costs, even when these are difficult to quantify.

Comparing our results to other literature, we see that estimates of the effectiveness of LFD testing vary widely, and are context dependent. In large populations (e.g. whole nations or regions), regular mass testing for prolonged periods is likely prohibitively expensive and so test, trace and isolate (TTI) strategies are more feasible. In studies of TTI, timing and fast turnaround of results is key \cite{fyles_2021,grassly_comparison_2020}, overall efficacy is lower than predicted here due to the targeted nature of testing (and the inherent `leakiness' of tracing contacts), however it is much more efficient than mass testing, particularly when incidence is low. Even without contact tracing, other `targeted' testing strategies, while not as effective as mass testing, can reduce incidence significantly \cite{gharouni_testing_2022} for a lower cost. Similarly, surveillance testing, of a combination of symptomatic and non-symptomatic individuals, is an efficient way to reduce the importation of new cases and local outbreaks \cite{lovell-read_interventions_2021}.  Focusing on mass LFD testing, as studied here, we predict a greater impact than the model \cite{hellewell_2021} and more similar to the model (although measured differently here) in \cite{ferretti_2021}  as we also use a viral load based model. Models fitted to real data in secondary schools suggests that twice weekly LFD testing would have reduced the school reproduction number by $\sim 40\%$ (if adherence reached 100\%), which is less effective than the 60\%--80\% (figure \ref{sec:main_results}) estimate provided here. As shown in table \ref{tab:sens_res1}, uncertainty in a number of parameters could explain this difference. The simplifying assumptions in this model are also likely to result in an over-estimate of effectiveness. For example, testing behaviour could be correlated with contact behaviour \cite{berrig_2022} and could provide false reassurance to those who are `paucisymptomatic' which would greatly reduce its benefit \cite{skittrall_sars-cov-2_2021} for the population as a whole. Similarly, infectiousness or testing behaviour may be correlated with symptoms, which could also skew these predictions depending on symptomatic isolation behaviours. Therefore, there is a need for integration of models of behaviour and engagement with testing policies into testing models to better predict its impact. 

There are other limitations of the models used in this study which need to be highlighted in order to interpret the results. First, the RNA viral load, testing and infectiousness data all pre-dates the emergence of the omicron variant (BA.1 lineages), which are characterised by higher reproduction numbers, shorter serial intervals, and less severe outcomes \cite{tanaka_2022,backer_2022,del_aguila-mejia_2022}. The shorter incubation period means repeat asymptomatic testing for omicron is likely to be less effective than predicted here, especially for PCR testing with a high turnaround time. On the other hand, if more people asymptomatically carry omicron \cite{garrett_2022}, then this will increase testing impact. Second, the relationships used to relate RNA viral load, infectiousness, and test positive probability are not representative of the mechanistic relationships between these quantities. Therefore, the test sensitivity relationship used will likely marginally overestimate the impact of very frequent testing (e.g. daily testing) since it does not take into account possible interdependence of subsequent test results (except for the correlation with RNA viral load). Other determinants can effect sensitivity and some studies suggest culture positive probability is a better indicator of LFD positive-probability than RNA viral load \cite{kirby_2021,pekosz_2021,pickering_2021,killingley_2022}. Similarly, while ``infectious virus shed'' is undoubtedly a factor in infectiousness, it is not the only determinant (as is assumed here). Other determinants of infectiousness (independent of contact rate) such as symptomatology, mode of contact, etc. mean that the relationship between viral load and infectiousness measured in contact tracing and household transmission studies can be much less sharp than used here (e.g. \cite{lee_2022}), although as discussed in \cite{ferretti_2021} both sharp and shallow relationships are plausible depending on the dataset used and different infectiousness profiles can change the relative impact of testing and symptomatic isolation \cite{hart_high_2021}. The sensitivity analysis presented in \ref{app:sens_analysis} shows that decreasing the parameter $h$ (which results in a less sharp relationship between viral load and infectiousness as well as a broader infectiousness profile, see figure B.1) significantly decreases the impact of testing. This change essentially increases the proportion of infectiousness that occurs before an individual is likely to test positive and isolate. Therefore, it is important to compare multiple different models starting with different sets of reasonable assumptions to generate predictions that inform policy and so models based empirical measures of infectiousness or different within host models will be a useful area of future research. Finally, we have not carried through the results on testing policies to their implications on epidemiological outcomes, such as hospitalisations and deaths, which would be required to perform a full cost-benefit analysis of different testing outcomes. 

In conclusion, repeat asymptomatic testing with LFDs appears to be an effective way to control transmission of SARS-CoV-2 in the workplace, with the important caveat that high levels of adherence to testing policy is likely more important than the exact testing regime implemented. Specificity of the particular tests being used must be taken into consideration for these policies, as even tests with high specificity can result in the same number of false positives as true positives when prevalence is low. The code used for the calculation of $\Delta$IP \cite{whitfield_model_2022-1} and the workplace simulations \cite{whitfield_model_2022} is available open-source. As we have shown, the detailed model of $\Delta$IP developed here can be used to simulate both the population-level change in effective infectiousness due to a change in testing policy, but also the individual-level effect. Direct interpretations of $\Delta$IP should be made with caution because they only quantify the personal reduction in transmission risk. While testing can reduce both ingress into and transmission within the workplace, repeated ingress and internal transmission could still result in a high proportion of individuals becoming infected (albeit at with a slower growth rate than the no testing case) even with testing interventions present and functional, depending on community prevalence and the length of time for which this prevalence is sustained. Nonetheless, calculations of $\Delta$IP have the potential to be used in existing epidemiological simulations to project the impact of testing policies without having to simulate the testing an quarantine explicitly, simply as a scale factor on the individual-level or population-level infectiousness parameters, depending on the model being used. 

\section*{Funding sources}

This work was supported by the JUNIPER modelling consortium (grant number MR/V038613/1) and the UK Research and Innovation (UKRI) and National Institute 561 for Health Research (NIHR) COVID-19 Rapid Response call, Grant Ref: MC\_PC\_19083. 

\section*{Acknowledgements}

We would like to acknowledge the help and support of colleagues Hua Wei, Sarah Daniels, Yang Han, Martie van Tongeren, David Denning, Martyn Regan, and Arpana Verma at the University of Manchester. Additionally we would like thank the Social Care Working Group (a subcommittee of SAGE) for their valuable feedback on this work. In particular we acknowledge several the insights gained from fruitful discussions with Nick Warren as part of his work for the Health and Safety Executive.

\section*{Author Contributions}

CAW and IH contributed to the model development, interpretation of the results and drafting and revising of the manuscript. CAW performed the data analysis and simulation design. Authors named as part of the ``University of Manchester COVID-19 Modelling Group'' contributed to the generation and discussion of ideas that fed informed this manuscript, but were not directly involved in the writing of it.

\bibliography{My_Library.bib}
\bibliographystyle{elsarticle-num.bst}

\appendix

\setcounter{figure}{0} 
\setcounter{table}{0} 

\section{Data and Simulation methods}
\label{app:sim_methods}

\subsection{Parameter values}

Supplementary table S1 gives the parameter values used in the models of viral load, infectiousness and test sensitivity as described in sections \ref{sec:viral_load_param}, \ref{sec:infectiousness_param}, and \ref{sec:test_sens} derived from sources \cite{peto_2021,ke_2021,smith_2020,nhs_test_and_trace_dual-technology_2021,overton_2020,pouwels_2021}. 

\subsection{Calculation of Infectious Potential}
\label{app:IP_calc}

To calculate IP for each individual we discretise equations \eqref{Rnum}, \eqref{VLmodel} and \eqref{infectiousness}. We choose a time-step of 1-day for computational efficiency and because this is the shortest time between tests that we consider. In practice, this means we assume that the viral load on day $t \in \mathbb{Z}$ is given by $V(t)$ for the whole day. Since the viral load can actually vary quickly, and therefore the infectiousness can vary between the start and end of a day, we account for this by discretise equation \eqref{infectiousness} as follows
\begin{align}
    J_t^{(k)} = \int_{t-0.5}^{t+0.5} J_k[V_k(t)] \textrm{d} t
\end{align}
where the integral is computed analytically using equations \eqref{VLmodel} and \eqref{infectiousness}. Thus, the infectiousness on day $t$ is given by the average infectiousness over the 24-hour period. This means that the integral in the calculation of IP (in equation \eqref{Rnum}) is discretised as
\begin{align}
    \textrm{IP}_k \approx \frac{1}{\langle{\tau}_{\rm inf}\rangle}\sum_{t=0}^{\tau_{\rm inf}^{(k)}} \left(1 - I_t^{(k)}\right) J_t^{(k)}
\end{align}
where $I_t^{(k)} = 0$ if individual $k$ is at work that day, and $I_t^{(k)} = 1$ if not. The day $t_{\rm max}^{(k)}$ is the last day for which an individual has a viral load exceeding $V_{\rm cut}$.

\begin{figure}[ht]
    \centering
    \includegraphics[width=0.6\textwidth]{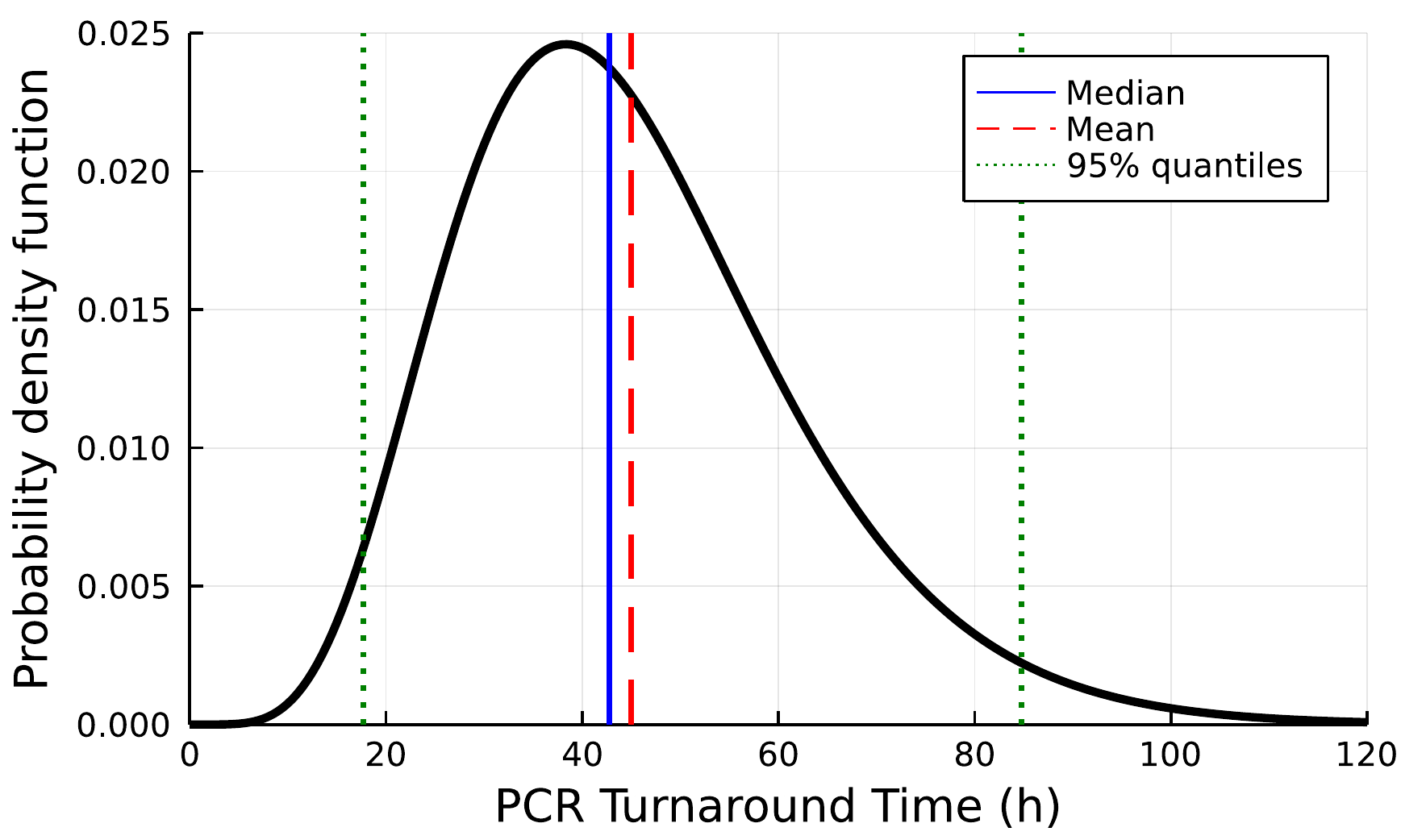}
    \caption{Probability distribution of PCR turnaround times used in this paper in hours. Vertical lines show the median, mean and 95\% central interval of this distribution, as labelled. This distribution was created to imitate data collected by NHS Test-and-Trace in October 2021, based on weekly PCR testing of staff working in high-risk public sector jobs. The times quotes are measured from the time the test was taken (at work) until the result was received (electronically) by the member of staff.}
    \label{fig:PCR_TaTs}
\end{figure}

To model isolation, test results and symptom isolation are drawn with the relevant probabilities for individual $k$ and if any trigger an isolation, the earliest isolation day becomes $t_{\rm isol^{(k)}}$. Note that symptomatic isolation is assumed to begin on the nearest whole number day to the randomly drawn symptom onset time. Similarly, for positive PCRs, isolation begins on the nearest whole number day from the test result (see figure \ref{fig:PCR_TaTs} for a summary of the turnaround times used). For positive LFDs, people isolate on the day they perform their test (so it is assumed to be taken at the start of the day, before any workplace exposure). Once $t_{\rm isol}^{(k)}$ has been determined for an individual, we set $I_t = 1$ for $t_{\rm isol} \leq < t_{\rm isol} + \tau_{\rm isol}$ and re-calculate their IP. 

To calculate $\langle \tau_{\rm inf} \rangle$ we generated $5 \times 10^6$ trajectories and calculated the average number of days for which individuals had a viral load $V_t > V_{\rm cut}$, i.e. the period of time they could test positive via PCR. These values are therefore different for the two viral load models, and are given in Supplementary table S1. The code used to perform all of these calculations is available at \cite{whitfield_model_2022-1}.



\subsection{Workplace outbreak simulations}
\label{app:wp_sims}

We use the same Julia program to simulate both workplace transmission scenarios outlined in section \ref{sec:workplace_models} \cite{whitfield_model_2022}. The simulations proceeds as follows.

At initialisation, the following model features are generated
\begin{enumerate}
    \item Agents are assigned roles and shift patterns
    \begin{itemize}
        \item All staff have the same role and boolean shift pattern. Each is drawn a random permutation number from 1-14 to determine when their shift pattern starts. 
        \item In the two-component model, all patients are also assigned a nominal ``shift pattern'', however this has value `true' for every day.
    \end{itemize}
    \item If there is a testing regime for staff:
    \begin{itemize}
        \item Staff are selected at random with probability $P_{\rm not}$ to be `non-testers'.
        \item Testing staff are assigned a boolean testing pattern which has the same start day as their shift pattern. 
        \item For all days labelled as a testing day, each is changed to a non-testing day with probability $P_{\rm miss}$.
    \end{itemize}
    \item  An index case is chosen at random and infected. Upon infection, an agent is assigned the following:
    \begin{itemize}
        \item Viral load and infectiousness trajectories (equations \eqref{VLmodel} and \eqref{infectiousness}).
        \item Symptom onset time.
        \item Boolean adherence to symptomatic isolation (true with probability $p_{\rm isol}$).
        \item If testing: a test positive probability trajectory (equation \eqref{PCRhigh} or Supplementary table S1).
    \end{itemize}
\end{enumerate}

The main simulation loop is executed for each day of the simulation, and proceeds as follows:
\begin{enumerate}
    \item Update infectious state of all individuals moving any to 'Recovered' status who have reached the end of their infectious period.
    \item Perform testing for all agents testing that day. For all positive tests generate an isolation time from the current day as $\lfloor \tau_d + u_{01}\rceil$ where $u_{01} \sim U(0,1)$ is a number uniformly distributed beteween 0 and 1, and $\lfloor .\rceil$ indicates rounding to the nearest integer (to simulate tests being performed before or after shifts, at random).
    \item Update isolation status for any who are due to isolate on this day.
    \item Identify all agents `on shift' on this day.
    \item Generate all workplace contacts of infectious agents:
    \begin{itemize}
        \item For each infectious agent with role $k$, generate all contacts with each job role $m$ by selecting from those on shift with probability $\textbf{P}^{(c)}_{k,m}$.
        \item Calculate the probability that each contact results in infection using the expression in equation \eqref{p_inf}.
    \end{itemize}
    \item Generate all successful workplace infection events at random with the assigned probabilities.
    \item For any infectees that are subject to more than one successful infection event, select the recorded infection event at random.
    \item Record all infection events, and for every individual infected change their status to 'infected' and their infection time to the current day. Their susceptibility is set to 0.
    \item Increment the day and return to step 1 unless the maximum number of days has been simulated or if no infectious agents remain in the simulation.
\end{enumerate}

\setcounter{figure}{0} 
\setcounter{table}{0} 

\section{Sensitivity Analysis}
\label{app:sens_analysis}

\subsection{Method}
\label{app:sa_method}

\begin{figure}[ht]
    \centering
    \includegraphics[width=\textwidth]{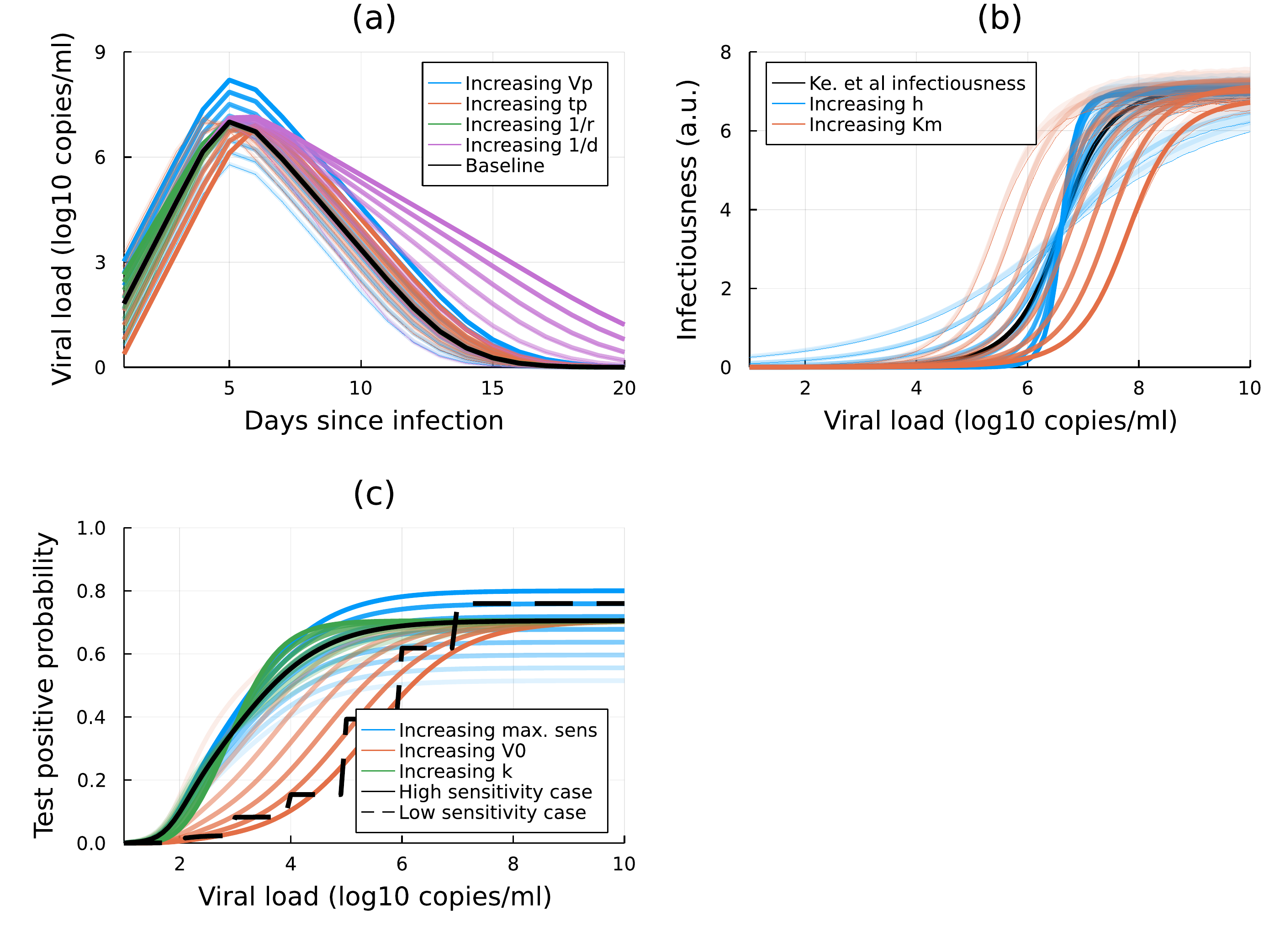}
    \caption{Visualisation of the parameter ranges used in the sensitivity analysis. In these figures, parameters are varied independently to show their individual influence across all of the values they take in the sensitivity analysis. Darker shading indicates higher values and the colour indicates the parameter that has been varied, as labelled. The shaded area around the curves shows 95\% confidence intervals in the mean, estimated using 1000 bootstrapping samples. (a) Population mean viral load trajectories (Ke et al. model -- black line) while varying the peak viral load $V_p$, peak time $t_p$, inverse growth rate $1/r$, and inverse decay rate $1/d$ parameters. (b) Population mean infectiousness relationships (Ke et al. model -- black line) while varying the slope $h$ and threshold viral load $K_m$ parameters. (c) Test-positive probability relationships (`high' and `low' sensitivity models shown by the solid and dashed black lines respectively) while varying the maximum sensitivity $\lambda$, the threshold viral load $V_{50}^{(l)}$, and slope $s_l$ parameters.}
    \label{fig:sens_vis}
\end{figure}

To estimate the sensitivity of the testing model to various parameter assumptions, we use an ``Elementary Effects'' approach \cite{satelli_elementary_2007} for the main 11 model parameters used for the LFD testing model. The prior distributions for these parameters have not been possible to estimate, given that most of them only come from a single source and only some of them have been the subject of meta-analyses. Therefore, using the information available we have set plausible ranges for the parameters we test in table \ref{tab:sens_params}, and visualised their effects on model inputs in figure \ref{fig:sens_vis}. 
\begin{longtable}{| m{3cm}<{\centering}
                  | m{2.2cm}<{\centering} | m{2cm}<{\centering} | m{5cm}<{\centering} |}
\hline Parameter & Distribution & Range & Literature values \\ \hline
Median peak VL $V_p$ & U$[\log(V_p)]$ & $10^{6.4}$ -- $10^{8.8}$ copies/ml  & $10^{5.6}$ -- $10^{7.0}$ \cite{ferretti_2021}, $\sim 10^{7.5}$ \cite{kissler_2021} $\sim 10^{7.6}$ \cite{ke_2021}, $\sim 10^{8.0}$ \cite{singanayagam_community_2022}, $\sim 10^{8.9}$ \cite{killingley_2022} . \\ \hline
Median peak VL time $t_p$ & U$[t_p]$ & 3.0 -- 5.0 days & 3.2 \cite{kissler_2021}, 4.0 \cite{ke_2021}, 5  \cite{killingley_2022}\\ \hline
Median VL inv. growth $1/r$ & U$(1/r)$ &  0.25 -- 0.35 days & 0.17 -- 0.23 \cite{ferretti_2021}, $\sim$0.25 based on \cite{killingley_2022},  0.29 \cite{kissler_2021}, 0.3 \cite{ke_2021}\\ \hline
Median VL inv. decay $1/d$ & U$(1/d)$ & 0.41 -- 1.0 days & Biased towards longer shedding durations than used here \cite{he_temporal_2020,singanayagam_community_2022,cevik_sars-cov-2_2021,killingley_2022}\\ \hline
Median inf. sigmoidal slope $h$ & U$(h)$ & 0.27 -- 3.0 & Lower (not quantified) \cite{lee_2022, marc_quantifying_2021, marks_transmission_2021}. Similar/higher (not quantified) \cite{goyal_viral_2021, ferretti_2021}\\ \hline
Inf. scale param. $K_m$ & U$[\log(K_m)]$ & $10^{5.4}$ -- $10^{7.8}$ copies/ml& Lower \cite{lee_2022,marc_quantifying_2021}, Higher \cite{goyal_viral_2021}\\ \hline
LFD max. sens. $\lambda$ & U$(\lambda)$ & 0.54 -- 0.84 & Varying between two sources used and incorporating lower values \\ \hline
LFD sens. cutoff $V_{50}^{(l)}$ & U$[\log_{10}(V_{50}^{(l)})]$ & $10^{2.4}$ -- $10^{5.4}$ copies/ml & Varying between two sources used.\\ \hline
LFD sigmoidal slope $s_l$ & U$[\log(s_l)]$ & 0.67 -- 2.2 & Chosen to vary between 2 sources used.\\ \hline
Symp. prob. $P_{\rm symp}$ & U$(P_{\rm symp})$ & 0.20 -- 0.80 & Dependent on symptomatic isolation criteria vaccination status.\\ \hline
Mean symp. onset $\mu_s$ & U$(\mu_s)$ & 3.3 -- 6.3 & Near to peak viral load \cite{walsh_sars-cov-2_2020, benefield_sars-cov-2_2020}. $\sim 5$ \cite{overton_2020} \\ \hline
\caption{List of parameters varies in the elementary effects sensitivity analysis. The `distribution' column shows the assumed parameter distribution that the parameters are evenly sampled across, where U denotes a uniform distribution. The `range' column gives the maximum and minimum values of these distributions used in the sensitivity analysis. The final column provides some justification for the ranges used. Note that only studies where nasal viral load data was collected in the incubation period was used to inform the ranges for peak viral load, timing and growth rate parameters.}
\label{tab:sens_params}      
\end{longtable}

The Elementary Effects method was performed as follows. The chosen (uniform) prior for each parameter $k\in\{1,\ldots,11\}$ was split into $p = 8$ equal quantiles. We will denote these quantiles for parameter $k$ by the vector $\mathbf{q}_k = [0,1/7,\ldots,1]$. Then, $r=50$ paths were drawn to sample the parameter quantile space. This was performed by first randomly drawing a starting point $\mathbf{q}^{(0)}$ (i.e. drawing a number from $\mathbf{q}_k$ for each of the 11 parameters) from the $8^{11}$ possible starting points in parameter-value space. Then each path consists of 12 points in this parameter space by taking steps of size $\Delta q$ in each of the 11 parameter dimensions in a random order. A step size of $\Delta q = p/(2(p-1))= 4/7$ was chosen to give equal probability sampling across the plausible parameter ranges. The step direction (positive or negative) was determined by the starting point (since $\Delta q > 1/2$ and so if $q_k^{(0)} > 0.5$ the step for the parameter $k$ has to be negative). The path forms the 11$\times$12 matrix $\mathsf{Q}$.

To improve the spread of these $r=50$ paths (i.e. ensure they are well separated in parameter space) we iteratively replaced paths with new random paths as follows
\begin{enumerate}
\item Calculate the distance $d_{ij} = \sum_{n_1=1}^{12}\sum_{n_2=1}^{12}\sqrt{\sum_{k=1}^{11} (Q_{n_1,k}^{(i)} - Q_{n_2,k}^{(j)})^2}$ for each pair of paths in the $r=50$ generated.
\item Calculate the path spread squared $D^2_{1,\ldots,r} = \sum_{i=1}^{r} \sum_{j=i+1}^{r} d_{ij}^2$.
\item Generate a new random path $\mathsf{Q}^{(r+1)}$. 
\item For each path $i=1,\ldots,r$, replace the path $i$ with the path $r+1$ and recalculate $D^2_{1,\ldots,i-1,i+1,\ldots,r+1}$. 
\item If any $D^2_{1,\ldots,i-1,i+1,\ldots,r+1} > D^2_{1,\ldots,r}$ for $i \in \{1,\ldots,50\}$, replace the path $i$ corresponding to the maximum value of $D^2_{1,\ldots,i-1,i+1,\ldots,r+1}$ with the path $r+1$ and return to step 3.
\end{enumerate}
We repeated this process 5000 times, at which point paths were being replaced infrequently (approx. every 50 iterations), suggesting that the paths were somewhat well spread.

For each step on each of the final $r=40$ paths, we ran two of the simulated scenarios considered in figure \ref{fig:scenarios}, namely the case with 2 LFDs per week and the case with Daily LFDs (with 100\% adherence to testing), using 1000 simulations per scenario. The outcome measure we use is $\Delta$IP which was calculated once again by simulating 10$^5$ realisations. 

The elementary effects for each parameter $k$ and path $i$ are then calculated as follows 
\begin{align}
    \mathrm{EE}_k(Q^{(i)}) = \frac{\Delta \mathrm{IP}(\mathbf{Q^{(i)}_{n+1}}) - \Delta \mathrm{IP}(\mathbf{Q^{(i)}_{n})}}{Q^{(i)}_{n+1,k} - Q^{(i)}_{n,k}},
\end{align}
where $n+1$ is the step in path $i$ where the parameter $k$ changes (i.e. $Q^{(i)}_{n+1,k} - Q^{(i)}_{n,k} = \pm \Delta q$). Then the summary statistics of the elementary effects for each parameter are defined as
\begin{align}
\mu_k^* &= \frac{1}{r} \sum_{i=1}^{r} \left|\mathrm{EE}_k(Q^{(i)})\right| \\
\mu_k &= \frac{1}{r} \sum_{i=1}^{r} \mathrm{EE}_k(Q^{(i)}) \\
\sigma^2 &= \frac{1}{r-1}\sum_{i=1}^{r} [\mathrm{EE}_k(Q^{(i)}) - \mu_k]^2
\end{align}
Finally, we repeated this process 10 times in total to estimate the uncertainty in $\mu^*_k$, $\mu_k$ and $\sigma_k$. 

\subsection{Results}
\label{app:sa_results}

The summary statistics for the sensitivity analysis are shown in tables \ref{tab:sens_res1} and \ref{tab:sens_res2}. For reference, from figure \ref{fig:scenarios}, we see that the baseline values of $\Delta$IP for these cases are in the range 0.6-0.9. Therefore, values of $\mu^* < 0.03$ correspond to a $<5$\% change in the result and can be treated as not having a significant effect on the predictions. In both cases this includes the inverse growth rate ($1/r$), peak viral load time $t_p$, mean symptom onset time $\mu_s$, and the infectiousness scale parameter $K_m$. Note that, we do not change the stipulation within the model that symptom onset time must occur within 2-days either side of peak viral load time, which may explain why neither of these parameters have a large effect (it has been highlighted elsewhere that the relevant timing of onset of infectiousness and symptoms has important implications for testing efficacy \cite{hart_high_2021}). 
\begin{longtable}{| m{4.5cm}<{\centering}
                  | m{2.5cm}<{\centering} | m{2.5cm}<{\centering} | m{2.5cm}<{\centering} |}
\hline \multirow{2}{*}{Parameter} & \multicolumn{3}{c|}{$\Delta$IP (2 LFDs per week)} \\ \cline{2-4}
& $\mu^*$ & $\mu$ & $\sigma$ \\ \hline
LFD max. sens. $\lambda$ & 0.200 $\pm$ 0.004 & 0.200 $\pm$ 0.004 & 0.073 $\pm$ 0.003\\ \hline 
LFD sens. cutoff $V_{50}^{(l)}$ & 0.191 $\pm$ 0.004 & -0.188 $\pm$ 0.005 & 0.110 $\pm$ 0.004\\ \hline 
Median peak VL $V_p$ & 0.136 $\pm$ 0.003 & 0.121 $\pm$ 0.004 & 0.111 $\pm$ 0.004\\ \hline 
Median inf. sigmoidal slope $h$ & 0.125 $\pm$ 0.006 & 0.113 $\pm$ 0.006 & 0.116 $\pm$ 0.006\\ \hline 
Symp. prob. $P_{\rm symp}$ & 0.096 $\pm$ 0.004 & -0.078 $\pm$ 0.004 & 0.091 $\pm$ 0.007\\ \hline 
Median VL inv. decay $1/d$ & 0.084 $\pm$ 0.005 & 0.017 $\pm$ 0.004 & 0.108 $\pm$ 0.007\\ \hline 
Inf. scale param. $K_m$ & 0.067 $\pm$ 0.004 & 0.039 $\pm$ 0.004 & 0.082 $\pm$ 0.007\\ \hline 
LFD sigmoidal slope $s_l$ & 0.067 $\pm$ 0.001 & 0.049 $\pm$ 0.004 & 0.069 $\pm$ 0.002\\ \hline 
Median VL inv. growth $1/r$ & 0.056 $\pm$ 0.002 & 0.032 $\pm$ 0.003 & 0.073 $\pm$ 0.005\\ \hline 
Median peak VL time $t_p$ & 0.047 $\pm$ 0.002 & -0.002 $\pm$ 0.003 & 0.069 $\pm$ 0.005\\ \hline 
Symp. onset $\mu_s$ & 0.044 $\pm$ 0.002 & -0.005 $\pm$ 0.002 & 0.061 $\pm$ 0.003\\ \hline 
\caption{Sensitivity of the $\Delta$IP measure to various model parameters in the case of testing with 2 LFDs per week (with 100\% adherence) vs. no testing. Results are sorted in descending order of $\mu^*$ value. Values given are the mean of 10 repeated sensitivity analyses $\pm$ the sample standard deviation (estimated by 100 bootstrap samples).}
\label{tab:sens_res1}
\end{longtable}

The same 4 parameters also have the largest effect on both cases simulated, namely the LFD maximum sensitivity $\lambda$, LFD sensitivity threshold $V_{50}^{(l)}$, the peak viral load $V_p$, and the slope parameter for the sigmoidal relationship between infectiousness and viral load $h$. In all of these cases, the effects appear to essentially always act in the same direction (i.e. $|\mu|_k \approx \mu^*_k$). The most obvious effects are the LFD sensitivity parameters, increasing $\lambda$ and decreasing $V_{50}^{(l)}$ improve the sensitivity of the LFD tests, and so these have a very large impact on $\Delta IP$. The increase in $\Delta IP$ is a similar effect, it essentially improves the sensitivity of the LFD tests since these are defined as a function of viral load. The most interesting effect is perhaps the parameter $h$. As shown in figure \ref{fig:sens_vis}(b) increasing $h$ makes the infectiousness vs. viral load relationship sharper, in effect this concentrates the infectious period around the time where testing is most sensitive, and thus increases $\Delta IP$. In the opposite case, where $h$ decreases, the infectiousness is more spread out, and people are more likely to be infectious before they test positive.

\begin{longtable}{| m{4.5cm}<{\centering}
                  | m{2.5cm}<{\centering} | m{2.5cm}<{\centering} | m{2.5cm}<{\centering} |}
\hline \multirow{2}{*}{Parameter} & \multicolumn{3}{c|}{$\Delta$IP (Daily LFDs)} \\ \cline{2-4}
& $\mu^*$ & $\mu$ & $\sigma$ \\ \hline
Median inf. sigmoidal slope $h$ & 0.170 $\pm$ 0.007 & 0.169 $\pm$ 0.007 & 0.132 $\pm$ 0.006\\ \hline 
LFD sens. cutoff $V_{50}^{(l)}$ & 0.151 $\pm$ 0.003 & -0.148 $\pm$ 0.003 & 0.111 $\pm$ 0.004\\ \hline 
LFD max. sens. $\lambda$ & 0.137 $\pm$ 0.003 & 0.137 $\pm$ 0.003 & 0.076 $\pm$ 0.009\\ \hline 
Median peak VL $V_p$ & 0.112 $\pm$ 0.003 & 0.068 $\pm$ 0.005 & 0.124 $\pm$ 0.004\\ \hline 
Symp. prob. $P_{\rm symp}$ & 0.089 $\pm$ 0.004 & -0.079 $\pm$ 0.004 & 0.087 $\pm$ 0.009\\ \hline 
Median VL inv. decay $1/d$ & 0.089 $\pm$ 0.004 & -0.035 $\pm$ 0.004 & 0.118 $\pm$ 0.006\\ \hline 
Inf. scale param. $K_m$ & 0.069 $\pm$ 0.002 & 0.056 $\pm$ 0.003 & 0.080 $\pm$ 0.006\\ \hline 
LFD sigmoidal slope $s_l$ & 0.047 $\pm$ 0.002 & 0.034 $\pm$ 0.002 & 0.054 $\pm$ 0.005\\ \hline 
Median peak VL time $t_p$ & 0.037 $\pm$ 0.002 & -0.010 $\pm$ 0.002 & 0.053 $\pm$ 0.005\\ \hline 
Median VL inv. growth $1/r$ & 0.035 $\pm$ 0.001 & 0.016 $\pm$ 0.002 & 0.047 $\pm$ 0.002\\ \hline 
Symp. onset $\mu_s$ & 0.032 $\pm$ 0.001 & 0.006 $\pm$ 0.002 & 0.046 $\pm$ 0.003\\ \hline 
\caption{Sensitivity of the $\Delta$IP measure to various model parameters in the case of daily testing with LFDs (with 100\% adherence) vs. no testing. Results are sorted in descending order of $\mu^*$ value. Values given are the mean of 10 repeated sensitivity analyses $\pm$ the sample standard deviation (estimated by 100 bootstrap samples).}
\label{tab:sens_res2}
\end{longtable}
\end{document}